\documentclass[12pt,preprint]{aastex}
\usepackage{subfigure}
\usepackage{natbib}

\usepackage{epsfig,amsmath,amssymb,amsfonts,mathrsfs,latexsym,graphicx,lscape}
\usepackage{hyperref}
\usepackage{times}
\usepackage{epsfig}
\usepackage{rotating}
\usepackage{sidecap}
\usepackage{longtable}

\shorttitle{Scaling relations of Nearby Clusters}
\shortauthors{zhaohh et al.}
\begin{document}

\title{A GOOD MASS PROXY FOR GALAXY CLUSTERS WITH XMM-Newton}

\author{Hai-Hui Zhao,  Shu-Mei Jia, Yong Chen,   Cheng-Kui Li , Li-Ming Song, Fei Xie}

\altaffiltext {} {Key Laboratory of Particle Astrophysics,
 Institute of High Energy Physics,
 Chinese Academy of Sciences,
 Beijing 100049; zhaohh@ihep.ac.cn}

\begin{abstract}

We use a sample of 39 galaxy clusters at redshift $z < 0.1$ observed by {\it XMM-Newton}
to investigate the relations between X-ray observables and total mass.
Based on central cooling time and  central temperature drop,
the clusters in this sample are divided into two groups: 25 cool core clusters
and 14 non-cool core clusters, respectively. We study the scaling relations
of $L_{\rm bol}$-$M_{\rm 500}$,  $M_{\rm 500}$-$T$,
$M_{\rm 500}$-$M_{\rm g}$ and  $M_{\rm 500}$-$Y_{\rm X}$, and also
the influences of cool core on these relations. The results show
that the $M_{500}$-$Y_{\rm X}$ relation has a slope close to the
standard self-similar value, has the smallest scatter and does not vary
with the cluster sample. Moreover, the $M_{\rm 500}$-$Y_{\rm X}$ relation
is not affected by the cool core. Thus, the parameter of $Y_{ X}$  may be
the best  mass indicator.

\end{abstract}


\section{Introduction}

Cluster mass is a key  parameter for us to explore the evolution of large-scale structure  of the
Universe and to test the cosmological models \citep{voit05, Vikhlinin09, Mantz10, Reichert11}.
The commonly-used method of  X-ray mass estimate is  assuming that the clusters are spherically
symmetric and the intracluster medium (ICM) is in hydrostatic equilibrium within the cluster
gravitational well \citep{Sarazin88}. But for some cases, like high-redshift, dynamically
unrelaxed  or  X-ray faint  clusters, the common X-ray mass estimate is not suitable. In
these cases, the cluster mass can be inferred from the relations between the total mass
and observables, such as luminosity ($L_{\rm X}$), X ray temperature ($T$), gas mass
($M_{\rm g}$), and  total thermal energy of the ICM ($Y_{X}$). These relationships are
predicted by the simple self-similar model of cluster formation, in which non-gravitational
effects are ignored and the energy emission is dominated by thermal bremsstrahlung. The relations
given by simulation and observation are different \citep{Kravtsov05,Nagai07,Chen07,Zhang08}, and these
relations also vary with  observable samples \citep{Arnaud07,Vikhlinin09,Chen07,Reichert11}.

Many simulations have  shown that the slope  of the $M$-$T$ relation was consistent with the self-similar
value of 1.5 \citep{Kravtsov05, Nagai07}. Chen et al. (2007) also got the same result, using an isothermal
model with  {\it ROSAT} and {\it ASCA} data. Investigating 10 relaxed clusters observed by {\it XMM-Newton},
Arnaud et al. (2005) found that the slope of this relation  was consistent with  the expectation for  hot
clusters ($T >$ 3.5 $\rm keV$),  but was steeper for the whole sample. Sanderson et al. (2003) researched this
relation using  66 clusters  in the  0.5-15 $\rm keV$ temperature range. They derived  a steeper slope of 1.84,
and there was no obvious difference between the high mass  and low mass parts.
It is unclear whether the $M$-$T$ relation is consistent with  the  value expected by the self-similar
model \citep{Chen07,Nagai07}; or this is true only for hot clusters \citep{Arnaud05}; or the slope is steeper
over the entire mass range \citep{Sanderson03}.

Simulations and observations   have shown that  the $M$-$M_{\rm g}$ relation was shallower  than the self-similar
prediction \citep{Arnaud05, Nagai07}. The discrepancy may be due  to the dependence of gas fraction  on the cluster mass.
Zhang et al. (2008) obtained a slope of $0.91 \pm 0.08$ for the $M$-$M_{\rm g}$ relation using a sample of
37 LoCuSS clusters from {\it XMM-Newton}  data.
They also derived a steeper slope of $0.97 \pm 0.08$, which was consistent with the self-similar expectation,
for the sub-sample of non-cool core clusters. The $M$-$M_{\rm g}$ relation is very complicated.
It is unclear whether the gas mass fraction  depends on the  mass \citep{Vikhlinin06,Giodini09}.
There are also differences in the baryon fraction between cool-core  systems and non-cool core
systems \citep{Eckert12}. The evolution of gas mass with cluster mass is not yet fully understood.

The $L_{\rm X}$-$M$ relation is  very important for the cosmological application.
 The $L_{\rm X}$-$M$ relations differ significantly from  different samples \citep{Maughan07, Chen07, Zhang08, Pratt09, Reichert11}.
Reichert et al. (2011) obtained  a slope of  $1.51 \pm 0.09$ by 14 literature samples.
Using the core excised luminosity,  Maughan et al. (2007) found that the slope was $1.63 \pm 0.08$.
 Chen et al. (2007)  showed  that the slope of $1.94\pm 0.15$ for the cool-core clusters (CCCs) agreed with
 that of $1.75 \pm 0.25$ for the non-cool-core clusters (NCCCs) within errors. But there was a obvious difference
 in the normalizations of the relations for the CCCs and the NCCCs. Zhang et al. (2008) found
 that the slope  ($2.01 \pm 0.74$) for the NCCCs  was shallower than that  ($2.32 \pm 0.70$) for the whole sample, while
 the errors were large.

$Y_{\rm X}$, the product of gas mass and X-ray temperature ($Y_{\rm X}$= $T$$ \cdot M_{\rm g}$), is another mass
proxy.  With the consideration of gas cooling and star formation, Kravtsov et al. (2006) found that the slope
of the $M$-$Y_{\rm X}$ relation  was in good agreement  with the self-similar prediction \citep{Nagai07}.
Arnaud et al. (2007) derived a slope of $0.548 \pm 0.027$ with a sample of 10 relaxed clusters, which was
slightly shallower than the expected value.

Different cluster samples provide different results for the scaling relations \citep{Arnaud05,Chen07,
Nagai07, Zhang08,Reichert11}. It is necessary to build a large sample covering a wide range of temperature
and including both CCCs and NCCCs observed by a telescope with high spatial resolution and sensitivity, such as {\it XMM-Newton} or {\it Chandra}.
Moreover, a better method which can give more accurate X-ray parameters is also essential.  To derive the cluster mass
through so-called ``scaling relations'', we should identify the best mass proxy. Ideally, a robust mass proxy should be
characterized by that: (1) simple power law relation and evolution that can be close to the prediction of the self-similar model.
(2) a low  scatter in cluster mass. (3) stable  relation does not vary with cluster sample.
(4) cool cores have little influences on the relations.

Numerical simulations show that the $Y_{\rm X}$ may be a good mass indicator with only $\approx$ 5\%-8\%
intrinsic scatter in $M$ \citep{Nagai07, Kravtsov06}, which is smaller than any other mass proxies
even in the presence of significant dynamical activity. Both simulations and observations show that the
intrinsic scatter in mass around the $M$-$T$ relation  is small ($\bigtriangleup$$M/M$ $\approx$ 0.10)
\citep{Arnaud05, Vikhlinin06, Nagai07}.  Arnaud et al. (2007) found that the scatter was the same for
the $M$-$Y_{\rm X}$ and $M$-$T$ relations. But the sample they used was small and only for
relaxed clusters. They could not study whether the scatter was insensitive to dynamical state \citep{Arnaud07, Kravtsov06}.
The $M_{g}$ is also as a low-scatter proxy for $M$ \citep{Stanek10, Okabe10, Fabjan11}, this choice is motivated by the fact
that $M_{g}$ can be measured independent of the dynamical state of the cluster. Some authors showed that the scatter in
the $M$-$M_{\rm g}$ relation was smaller than that for the $M$-$Y_{\rm X}$ relation \citep{Okabe10, Fabjan11}.
For the hot and massive clusters, Mantz et al. (2010) found that the intrinsic scatter in the center-excised $L_{X}$-$M$ relation ($<$10\%)
was smaller than in either the $M$-$T$ or $M$-$Y_{\rm X}$ relation (10-15\%).

In this paper, we investigate a flux-limited sample of 39 X-ray  nearby ($z < 0.1$) galaxy clusters based
on  {\it XMM-Newton} observations. This sample covers a wide range of temperature (2-9 $\rm keV$), and
 includes both CCCs and NCCCs. Using de-projecting method, we can derive
 accurate intra-cluster medium (ICM) temperature profiles and density distributions.
 The main goals of this work are:
 (1) to derive precise X-ray cluster parameters, e.g., $T$, $L_{\rm X}$, $M$ and $M_{g}$, then to present four mass
 scaling relations: $L_{\rm X}$-$M$,  $M$-$T$,  $M$-$M_{\rm g}$ and  $M$-$Y_{\rm X}$;
 (2) to investigate the influences of cool core on the mass scaling relations;
 (3) to compare these  scaling relations aiming to find which parameter is the best mass proxy.
Throughout this paper, the energy band we select is 0.5-10 keV. We use a cosmological model with
$\Omega_{\rm M}$ = 0.3, $\Omega_{\rm \Lambda}$ = 0.7 and $H_{0}$ = 70 km s$^{-1}$ Mpc$^{-1}$.
All uncertainties are in 68\% confidence level.

\section{Sample Selection}

Using a flux-limited ($f$ $\geq$ 1.0 $\times$ $10^{-11}$  {$\rm erg$ $\rm s^{-1}$ $\rm cm^{2}$})
method, we select a nearby ($z$$<$$0.1$) regular galaxy cluster sample  from
RASS \citep{Grandi99}, HIFLUGCS \citep{Thomas02}, REFLEX \citep{Bohringer04},
NORAS \citep{Bohringer00}, XBACs \citep{Ebeling96} and BCS \citep{Ebeling98}
catalogs.
Some clusters (e.g., A168,
A2634, A3395), which are too weak  to obtain their basic parameters,
have been excluded from our sample.  Because of the large angular size, we
exclude the Coma, A3526, FORNAX, Perseus, Ophiuchus and Virgo clusters.
Moreover, the clusters (e.g., A3562, A3266, A3667) with obvious
substructures are unsuitable for  detailed de-projected analysis and are excluded.
 At last, we select a  sample of 39 clusters and all these
clusters are available from {\it XMM-Newton} as listed in Table 1.
This sample covers a broad temperature range of 2-9 keV.

To investigate the morphological characters of our sample, we  calculate the
values of centroid shift $w$, defined as the standard deviation  between the X-ray
surface brightness peak and the centroid of the system \citep{Pratt09}:

\begin{equation}
\langle w \rangle = \left[ \frac{1}{N-1} \sum \left(\Delta_i - \langle \Delta \rangle \right)^2 \right]^{1/2} \times \frac{1}{r_{500}},
\end{equation}

\begin{figure}
\resizebox{\hsize}{!}{\includegraphics[height=3.5cm,width=5.5cm]
{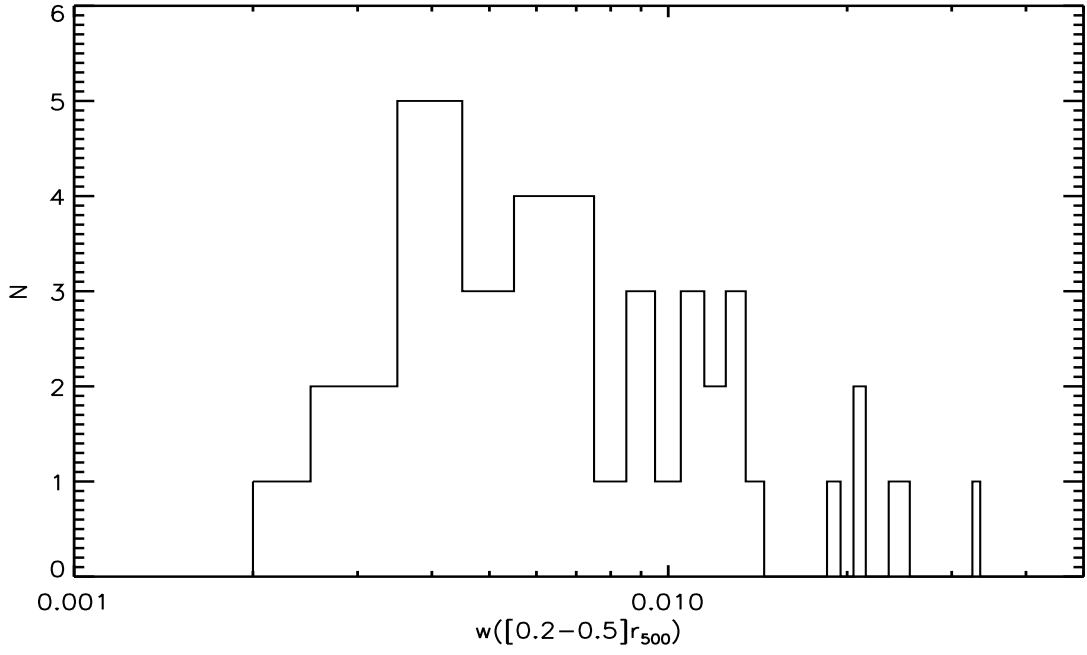}}
\caption[ ]{Histogram of central shift parameter $\langle w \rangle $, evaluated in the $[0.2-0.5]r_{500}$ aperture.}
\label{zhaohh:fig1}
\end{figure}

We divide the 2D projected image into several concentric apertures centered on the X-ray surface brightness peak.
The radii of apertures are $n\times 0.05\times r_{500}$ with n=4,5,6...10,
excluding the central regions to avoid bias associated with the bright central core.
We obtain the centroid of each aperture by determining the ``centre of mass" of the photon distribution.
$\Delta_i$ is the projected distance between the X-ray peak and the centroid in the $i$\,th aperture.
The distribution of $\langle w \rangle$ for our sample is shown in Fig.\ref{zhaohh:fig1}, which shows
that the upper limit of $\langle w \rangle$  is 0.04. Compared to the upper limit value of 0.1
in Pratt et al. (2009), our sample have a more regular morphology.

\section{Observations and Data Preparation}

The basic data reduction is done with Science Analysis System (SAS) 11.2.0.
In this paper, we consider the pn/EPIC data which are taken in Extended Full Frame  mode or Full Frame mode.
We only use the events with FLAG = 0, PATTERN $\leq$ 4,  the read out of time (OOT) effects  are also  corrected.

Since the X-ray flux of the cluster should be unchanged during the
observation period, we discard all the intervals with prominent flares and then select only those intervals
with count rates within 3$\sigma$ of the residual average count rate. The {\it XMM-Newton}
background can approximately be divided into two components. One is particle background, which
dominates at high energy and has little or no vignetting.
The other is Cosmic X-ray Background (CXB), which varies across the sky \citep{Snowden97},
more important at low energies, and shows significant vignetting.
We use the observations of $'$Lockman Hole$'$ (observation ID: 0147511801, hereafter LH) to  subtract these two
kinds of background. We estimate the ratio  of the particle background between the  cluster
and LH from the total count rate in the region $\theta \geq 10'$ in the high energy  band (12-14 keV), as
described in Pointecouteau et al. (2004).
We use the outer flat region of  the X-ray surface brightness distribution to monitor the residual background.
 At last, we apply a double-background subtraction method to correct for these two background components as used
 in Jia et al. (2004, 2006). The vignetting effects are also corrected.

All the galaxy clusters in this sample appear to be relaxed, therefore we assume the system structure to be spherically
symmetric. We extract the spectra from annular regions centered on the X-ray emission peak. In order to ensure the signal
to noise ratio, we use the criterion of $\sim$ 2000 net counts in 2-7 keV band per bin to determine the width of each
ring \citep{Zhang06, Zhang07}. The minimum width of the rings is set at 0.5$'$, which is wide enough for us to ignore
the Point Spread Function (PSF) effect of {\it XMM-Newton} EPIC, whose Full Width at Half Maximum (FWHM) is 6$''$ for pn.
Then, we can derive the de-projected spectra by subtracting all the contributions from the outer
regions (see Jia et al. 2004, 2006 for detailed calculation).

\section{X-Ray Properties}

\subsection{De-projected Temperature and Electron Density Profiles}
The spectral analysis is carried out by using XSPEC version 12.6.0. We fit the de-projected spectra with the absorbed Mekal model:

\begin{equation}
{\rm Model} = {\rm Wabs}(N_{\rm H})\times {\rm Mekal}(T,z,A,norm),
\end{equation}
in which Wabs is a photoelectric absorption model \citep{Morrison83}, Mekal is a  plasma emission
model \citep{Mewe85, Kaastra92}. The temperature $T$, metallicity $A$ and $norm$ (emission measure)
are free parameters. The redshift $z$ are fixed as in Table 1. For the majority of observations,
the absorption $N_{\rm H}$ are fixed parameters, in a few case (i.e., A478,  EXO 0422,
Hercules), $N_{\rm H}$ are not unique, we left them as a free parameter. After obtaining
the temperature of each shell, we can fit the radial de-projected temperature profile
by the following equation \citep{Xue04}:
\begin{equation}
{T(\rm r)} = {T_{\rm 0}}+\frac{A}{r/r_{\rm 0}}\exp(-\frac{(\ln r-\ln r_{\rm 0})^{2}}{\omega}),
\end{equation}
where $T_{\rm 0}$, $A$, $r_{\rm 0}$, and $\omega$ are free parameters.

For calculating the de-projected electron density profile, we divide the cluster into several
annular regions ($>$ 14, depending on count rate of the cluster) centered on the emission peak.
Then we use the de-projecting technique (see Jia et al. 2004, 2006 for detailed calculation)
to calculate the  photon counts in each shell. Since the de-projected temperature and abundance
profiles are known, we can estimate the normalization  `$norm (i)$' and its error in each shell.
Then we can  derive the de-projected electron density $n_{\rm e}$ of each region from Eq. (4).

\begin{equation}
 norm(i) = \frac{10^{-14}}{4\pi [D(1+z)]^{2}}\int n_{\rm e}n_{\rm H}dV,
\end{equation}
where $D$ is the angular size distance to the source in cm.
To obtain an acceptable fit for all clusters in this sample, we adopt a double-$\beta$ model to
fit the electron density profile \citep{Chen03}:
\begin{equation}
n_{\rm e}(r) = n_{\rm 01}[1+(\frac{r}{r_{c1}})^{2}]^{-\frac{3}{2}\beta_{1}}+ n_{\rm 02}[1+(\frac{r}{r_{c2}})^{2}]^{-\frac{3}{2}\beta_{2}},
\end{equation}
where $n_{\rm 01}$ and $n_{\rm 02}$ are  electron number density parameters, $\beta_{1}$ and
$\beta_{2}$ are the slope parameters, and $r_{\rm c1}$ and $r_{\rm c2}$ are
the core radii of the inner and outer components, respectively.

The primary parameters of all 39 galaxy clusters are given in Table 1.

\subsection{Mass Distribution}
Once we have obtained the de-projected radial profiles of electron density $n_{\rm e}(r)$
and temperature $T(r)$, together with the assumptions of hydrostatic equilibrium and spherical symmetry,
the gravitational mass of cluster within radius $r$ can be determined as \citep{Fabricant80}:
\begin{equation}
M( < r ) = -\frac{k_{\rm B}Tr^{2}}{G\mu m_{\rm p}}[\frac{d(\ln n_{\rm e})}{dr} + \frac{d(\ln T)}{dr}],
\end{equation}
where the mean molecular weight $\mu$
is assumed to 0.62. $k_{\rm B}$, $G$ and $m_{\rm p}$ are the Boltzman constant, the gravitational
constant, and the proton mass, respectively. The mass of  hot gas is calculated as:
\begin{equation}
M_{\rm g}( < r ) = 4 \pi \mu_{\rm e} m_{\rm p}\int n_{\rm e}(r)r^{2}dr,
\end{equation}
 where $\mu_{\rm e}$ is the mean molecular weight of the electrons. We calculate the total mass and
 gas mass within $r_{\rm 500}$, in which the mean gravitational mass density is equal to 500 times the
 critical density at the cluster redshift. The total masses, $M_{\rm 500}$, and gas masses, $M_{\rm g}$,
 within $r_{\rm 500}$ for all the clusters are listed in Table 2.

\subsection{Cooling Time}
The cooling time $t_{\rm cool}$ is a timescale during which the hot gas loses all of its thermal energy.
We calculate the cooling time of the gas as
\begin{equation}
t_{cool}=\frac{5}{2}\frac{n_{e}+n_{i}}{n_{e}}\frac{kT}{n_{H}\Lambda(A,T)},
\end{equation}
where $\Lambda (A, T)$ is the cooling function of the gas and we fix the metallicity at $A = 0.3\mbox{} Z_{\odot}$.
$n_{\rm e}$, $n_{\rm H}$, and $n_{\rm i}$  are the number densities of the
electrons, hydrogen, and ions, respectively. For the almost fully
ionized plasma in clusters, $n_{\rm e}$ = 1.2$n_{\rm H}$ and $n_{\rm i}$ = 1.1$n_{\rm H}$.
The central cooling time $t_{c}$ is derived from the central electron density $n_{e0}$ and
the central temperature $T_{0}$. $n_{e0}$ is the electron density at r=0.004$r_{500}$ \citep{Hudson10}
and  $T_{0}$ is the average temperature within 0-0.05$r_{500}$, where 0.05$r_{500}$
is the maximum value of the innermost annulus for the whole sample.

\section{Partition of CCCs and NCCCs}

Up to now, there are many methods used to distinguish CCCs from NCCCs, but it is unsure which is the best.
 CCCs is defined differently often based on a significant central temperature drop
\citep{Sanderson06, Arnaud07,Burn08}, short central cooling time  \citep{Bauer05, Ohara06, Donahue07},
significant classical mass deposition rate \citep{Chen07} or low central entropy
\citep{Hudson10}. Hudson et al. (2010) found that central cooling time was the best parameter
for low redshift clusters, and that cuspiness (defined as $\alpha=-\frac{d\log(n)}{d\log(r)}$)
was the best parameter for high redshift cluster.

We  define CCCs by two criteria, (1) the central cooling time, $t_{\rm c}$, is shorter
than $7.7{h_{\rm 70}}^{-1/2}$ Gyr \citep{Rafferty06}; (2) there is an obvious temperature
drop  compared with the peak temperature ($> 30\%$) towards the cluster center. Using this
two criteria, we divide our sample into NCCCs and CCCs, the fractions are 36\% and 64\%, respectively.

\begin{figure}
\resizebox{\hsize}{!}{\includegraphics[height=3.5cm,width=5.5cm]
{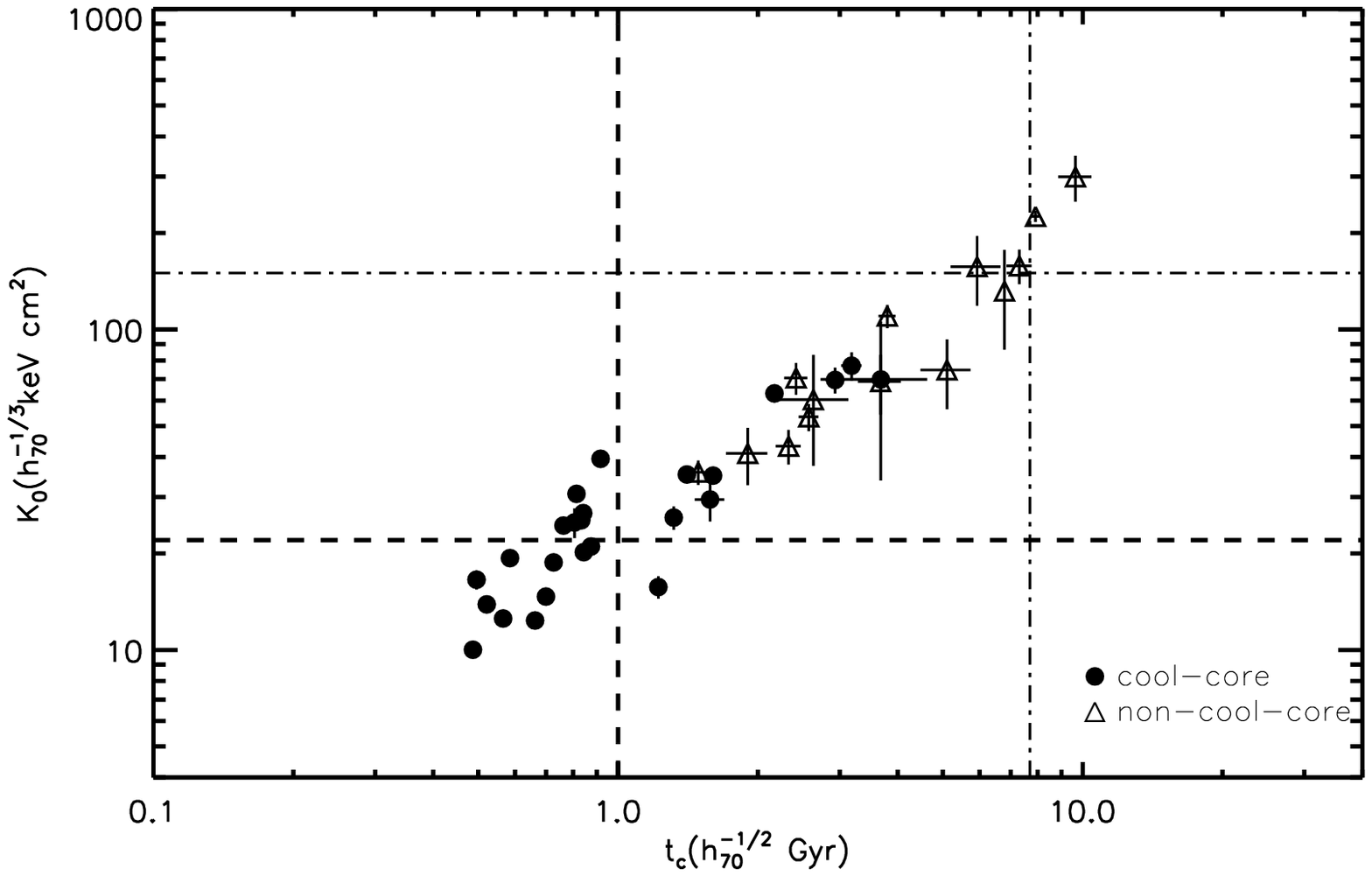}}
\caption[Central Entropy $(K_{\rm 0}$ versus Central Cooling Time $t_{\rm c}$ ]{This plot
shows central entropy ($K_{\rm 0}$) versus central cooling time $t_{\rm c}$. The $K_{0}$
can divide a sample at $\sim 22$ and $\sim$150 ${h_{\rm 70}}^{-1/3}$ $\rm keV$ $\rm cm^{2}$,
for the strong cool-core cluster (SCCC), weak cool-core cluster (WCCC), and NCCC \citep{Hudson10},
respectively. $t_{\rm c}$  can also divide a sample at 1.0 ${h_{70}}^{-1/2}$ Gyr and
7.7 ${h_{70}}^{-1/2}$ Gyr \citep{Vikhlinin07,Hudson10}. Combining $K_{\rm 0}$ and
$t_{\rm c}$,  the dash lines show the division between SCCC and WCCC, the dash-dot
lines show the division between  WCCC and NCCC. The filled circles represent the
pronounced CCCs, and the open triangles show NCCCs.}
\label{zhaohh:fig2}
\end{figure}

The central entropy, $K_{0}$ ($K_{0}=kT_{0}n_{e0}^{-2/3}$), is another  parameter to distinguish CCCs from NCCCs \citep{Hudson10}.
The $K_{0}$ can divide a sample at $\sim 22$ and $\sim$150 ${h_{\rm 70}}^{-1/3}$ $\rm keV$ $\rm cm^{2}$,
for the strong cool-core cluster (SCCC), weak cool-core cluster (WCCC), and NCCC \citep{Hudson10}, respectively.
$t_{\rm c}$  can also divide a sample at 1.0 ${h_{70}}^{-1/2}$ Gyr and 7.7 ${h_{70}}^{-1/2}$ Gyr \citep{Vikhlinin07,Hudson10}.
In Fig.\ref{zhaohh:fig2}, we compare the difference in sorting the sample using $K_{\rm 0}$ or $t_{\rm c}$.
 Combining $K_{\rm 0}$ and $t_{\rm c}$,  the dash lines show the division between SCCC and WCCC,
 the dash-dot lines show the division between  WCCC and NCCC.
The filled circles represent the pronounced CCCs, and the open triangles show NCCCs.
Based on Fig.\ref{zhaohh:fig2}, there is no much difference in sorting the sample using $K_{\rm 0}$ or $t_{\rm c}$.
With our definition of the CCC, we divide the WCCCs into CCCs and NCCCs.

%-----------------------------------------------------------------

%\clearpage
%------------------------------------------------------------------

\section{Self-similarity of the Scaled Profiles of the X-Ray Properties}

\subsection{Scaled Temperature Profiles}

\begin{figure}
\centerline{\hbox{\psfig{figure=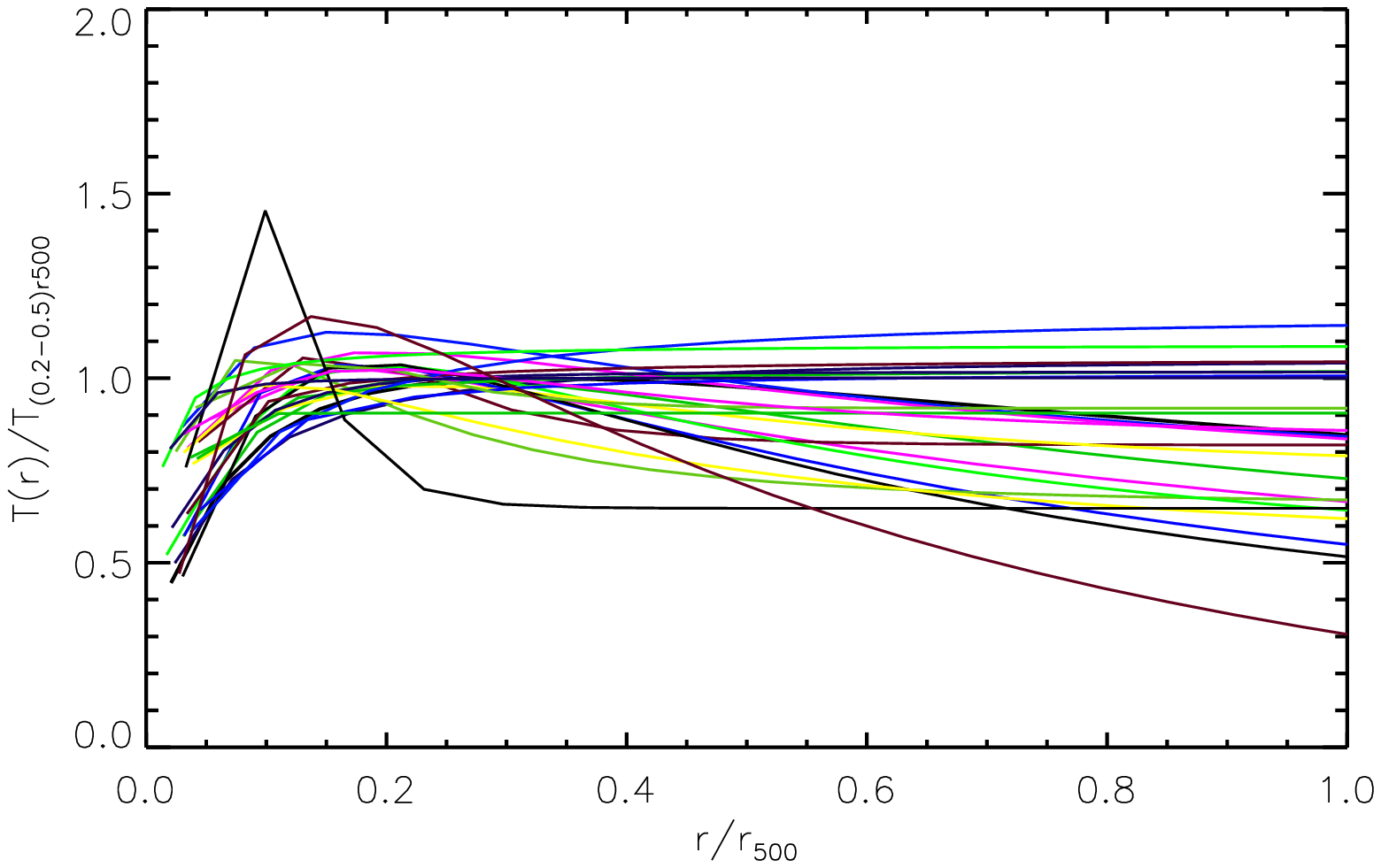,width=9.0cm,height=6.5cm}
\psfig{figure=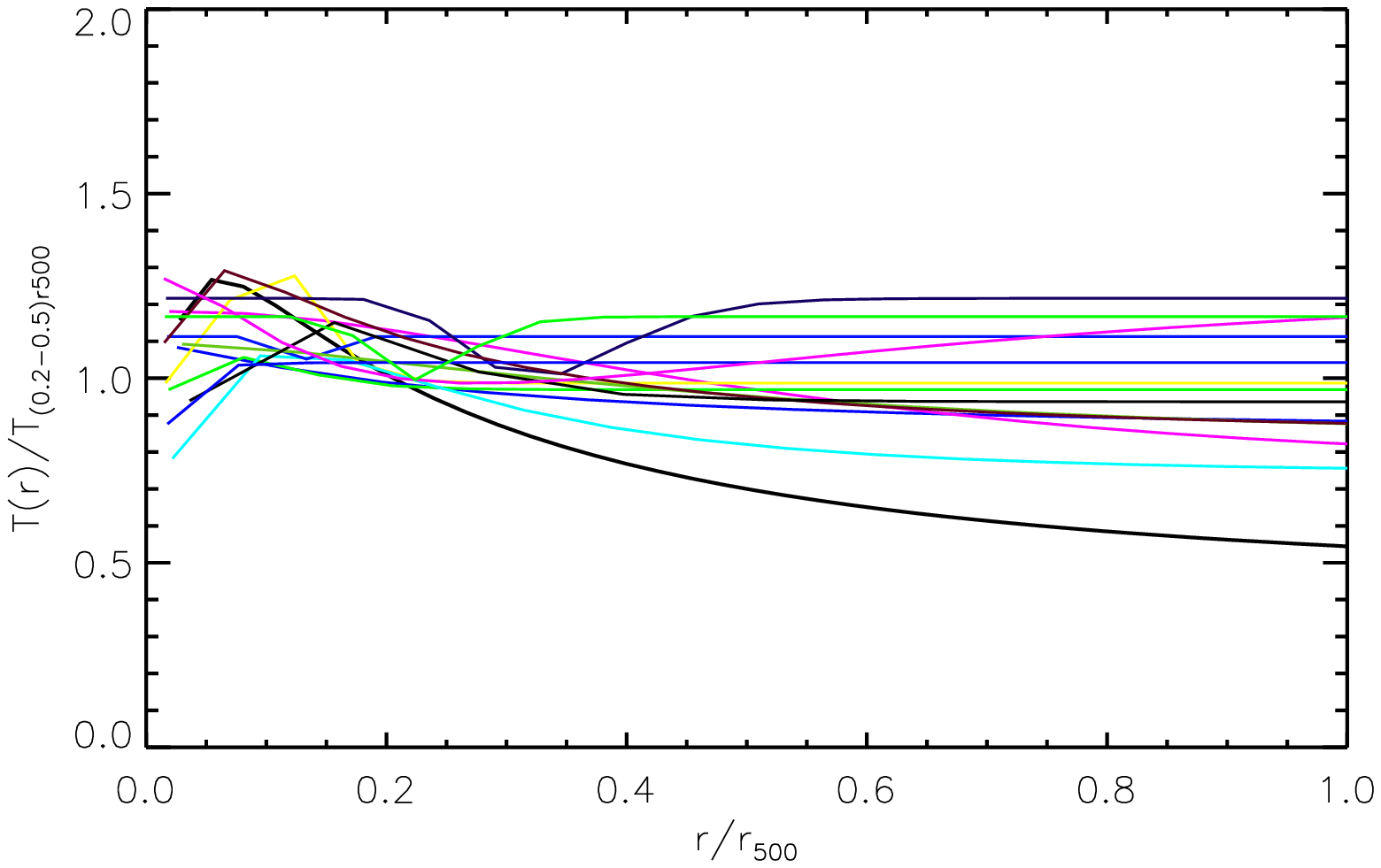,width=9.0cm,height=6.5cm}}}
\caption[Scaled radial temperature profiles of the CCCs]{Temperature profiles. Left panel: The temperature
profiles for  the CCC sub-sample. Right panel:  The temperature profiles for the NCCC sub-sample. We scale
the temperature profiles by $T_{(0.2-0.5)r_{\rm 500}}$ and $r_{\rm 500}$.}
\label{zhaohh:fig3}
\end{figure}

We derive the global temperature, $T_{(0.2-0.5)r_{\rm 500}}$, by the volume
average of the radial temperature profile limited to the radial range of
(0.2-0.5)$r_{\rm 500}$, as listed in Table 1. The temperatures within
0.2$r_{\rm 500}$ tend to show peculiarities linked to the cluster dynamical
history, which are mainly  affected by the cool cores \citep{Smith05}.
The upper boundary of 0.5$r_{\rm 500}$ is limited by the
quality of the spectral data. Thus, using  temperature within
(0.2-0.5)$r_{\rm 500}$, we can minimize the scatter in the X-ray scaling
relations and  reach a better agreement between the X-ray scaling relations for the CCCs and
NCCCs.   Fig.\ref{zhaohh:fig3} shows the temperature profiles for  the CCC sub-sample
and the NCCC sub-sample, respectively. There is a special cluster (Hercules), which has a rapid
fluctuation in  the temperature profile for  the CCC sub-sample, and  the unique distribution is also confirmed
by  the {\it Chandra} data \citep{Cavagnolo09, Li13}.

\subsection{X-ray Scaling Relations}
In the following, we investigate the relationships between several parameters.
We perform the relations with  logarithmic
values of the parameters in the form:

\begin{equation}
\log_{10}(Y) = A + B \cdot \log_{10}(X),
\end{equation}
where $X$ and $Y$ represent the variables, $A$ and $B$ are the two free parameters
to be estimated. We firstly create the histogram of residuals from  the best fitting
relation in the log space , and then, the raw scatter can
be obtained by a Gaussian fitting to the histogram. The
intrinsic scatter is calculated as \citep{Morandi07}:
\begin{equation}
S = \left[ \sum_{j}
\left(\left(\log_{10} (Y_{j}) -A -B \log_{10} (X_{j}) \right)^2
-\epsilon_{\log_{10} (Y_{j})}^2\right) / (N-2) \right]^{1/2}\ ,
\end{equation}
where $\epsilon_{\log_{10} (Y_j)} = \epsilon_{Y_{j}}/(Y_j \ln 10)$, with
$\epsilon_{Y_j}$ being the statistical error of the measurement $Y_j$,
and  $N$ is the total number of data.

\subsection{ The $M_{\rm 500}$-$T$ Relation}

\begin{figure}
\centerline{\hbox{\psfig{figure=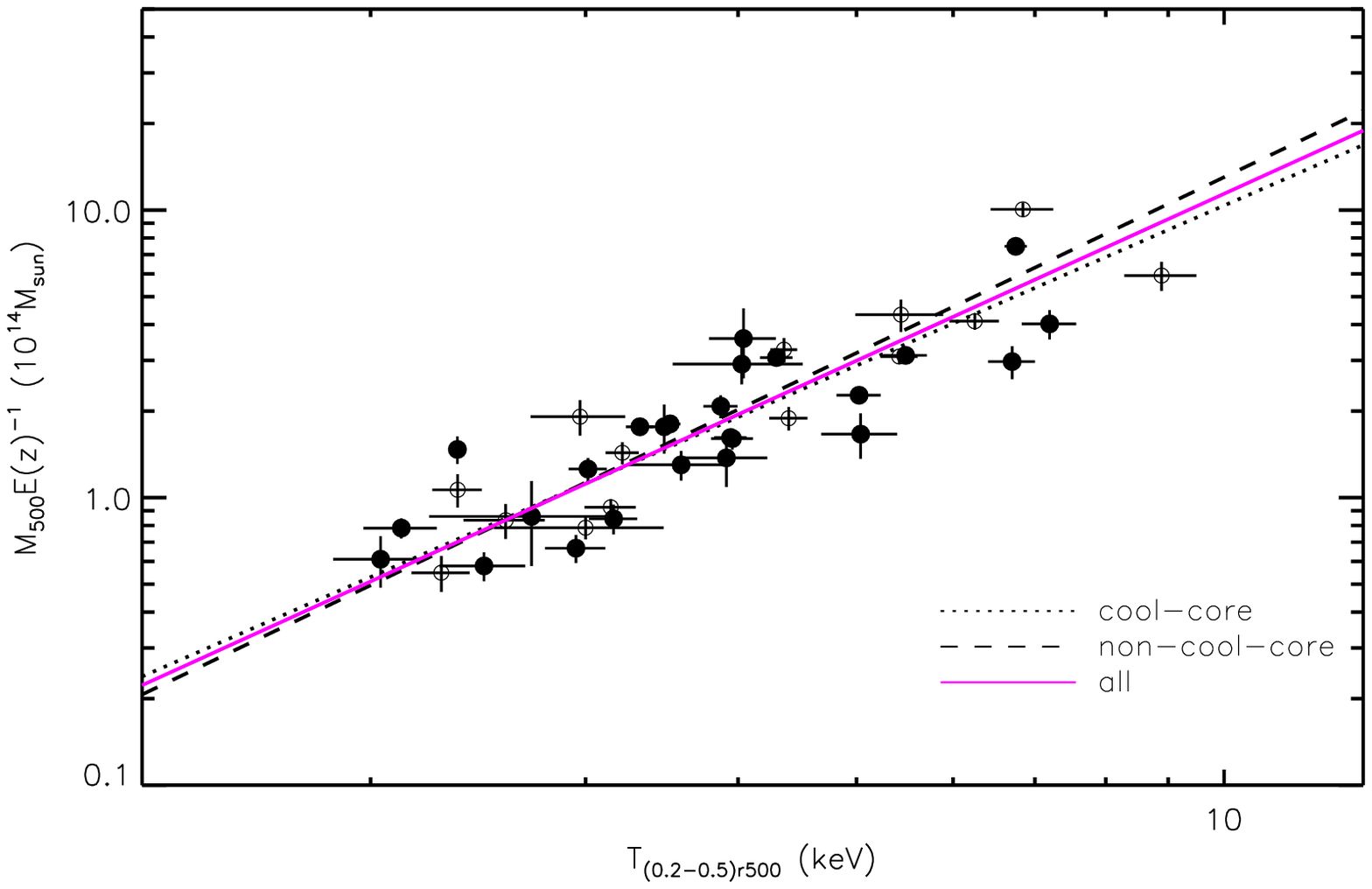,width=9.0cm,height=6.5cm}
\psfig{figure=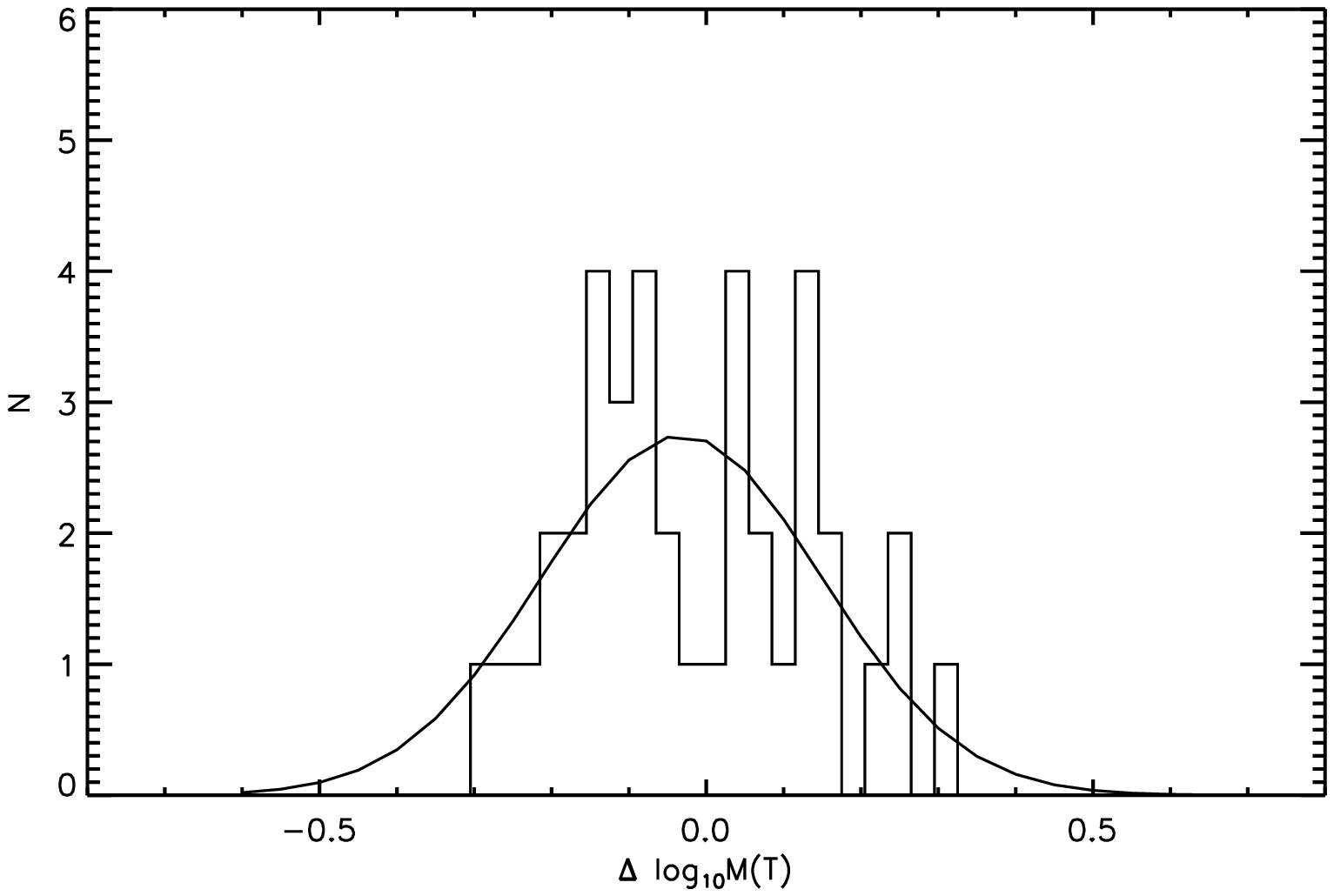,width=9.0cm,height=6.5cm}}}
\caption {Left panel: Total mass versus temperature diagram. In the following, the filled circles
and open circles represent the CCCs and  NCCCs, respectively; $E^{2}(z)=\Omega_{\rm m}(1+z)^{3}+\Omega_{\rm \Lambda}$
describes the evolution of the Hubble constant for a flat universe;  The red solid line shows the fit
for all the data and the dashed and dotted lines are those for the NCCCs and CCCs, respectively. Right
panel: Histogram of the residuals from the $M_{\rm 500}$-$T$ relation in log space. This raw scatter is
well described by a Gaussian fit  of $\sigma_{M(T)}$ = 0.18 dex.}
\label{zhaohh:fig4}
\end{figure}

%\clearpage

It is well known that different regression methods may give different slopes even at the  same population
level \citep{Akritas96}. Therefore, it is important to choose the most suitable method for different
data. In $M_{\rm 500}$-$T$ relation, we compare four kinds of BCES regression methods (BCES(Y$|$X),
BCES(X$|$Y), BCES Bisector, BCES Orthogonal), all of which take into account both  intrinsic scatter
and the presence of errors on both variables \citep{Akritas96}. The best power-law fits given by  these
four methods are shown in Table 3. The slopes from these four methods are consistent within $1\sigma$
errors. In order to compare with previous results, we use the  BCES Bisector
method. The results of power-law fits in the BCES Bisector method for all the relations discussed
in  section 6 are summarized in Table 4.

Fig.\ref{zhaohh:fig4} shows the  $M_{\rm 500}$-$T$ relation for our sample.
Considering the whole sample, our best-fit slope $1.94\pm 0.17$ is  higher than  the self-similar
predictive value, 1.5. The slope is consistent with the result in Mantz
et al. (2010), $2.08\pm 0.08$ for a 238  galaxy cluster sample  observed by  {\it Chandra} or {\it ROSAT}.
Reichert et al. (2011) derived a  slope of $1.76 \pm 0.08$ from the 14 literature samples.
Our slope  is also consistent with the result of Sanderson et al. (2003), $1.84\pm 0.01$
for the $M_{\rm 200}$-$T$ relation using {\it ASCA} temperature profiles. Previous studies have suggested
that the slopes of the  $M_{\rm 200}$-$T$ relation for the high mass and low mass parts may be different
\citep{Finoguenov01,Dos07}. The cross-over temperature between the two parts is typically $\sim3.0$ keV
\citep{Finoguenov01}.
Sanderson et al. (2003) investigated a  sample, which included 66 clusters with a broad range
of temperature (0.5-15 keV), and found  no obvious  break in the $M_{\rm 200}$-$T$ relation at $\sim$ 3.0 keV.
We also obtain a similar slope of $1.99 \pm 0.30$ for a sub-sample with $T > 3.0$ keV from
our sample. Thus, our slope is steeper than the self-similar prediction, and it does not vary with the temperature
(or mass) range of the sample.

Fixing  the slope  to  2.08,  we find that the normalization of our relation is lower
by $ \sim 28\%$ than that  in  Mantz et al. (2010).
With the slope fixed to 1.76,  the normalization for our sample is lower
than that in  Reichert et al. (2011) by $ \sim 27\%$ .

Moreover, the slopes and the normalizations of the $M_{\rm 500}$-$T$ relation for the CCCs and NCCCs
are consistent within errors as listed in Table 4. This indicates that no evident influence of cool core is
found on this relation. This agrees with the result obtained by Chen et al. (2007), who considered
a 88 cluster sample based on {\it ASCA} and {\it ROSAT} observations.

The right panel of Fig.\ref{zhaohh:fig4} shows a histogram of the log space residuals from the best fitting
$M_{\rm 500}$-$T$ relation for the whole sample. The residuals  are corresponding  to the value of
vertical  distances to the best fitting line \citep{Pratt09}. The raw scatter of the $M_{\rm 500}$-$T$
relation is well described by a Gaussian fit in log space with $\sigma_{M(T)}$ = 0.18 dex.
The intrinsic scatter in $M_{\rm 500}$ around the $M_{\rm 500}$-$T$ relation is 0.14 dex.

\subsection{The $M_{\rm 500}$-$M_{\rm g}$ Relation}

\begin{figure}
\centerline{\hbox{\psfig{figure=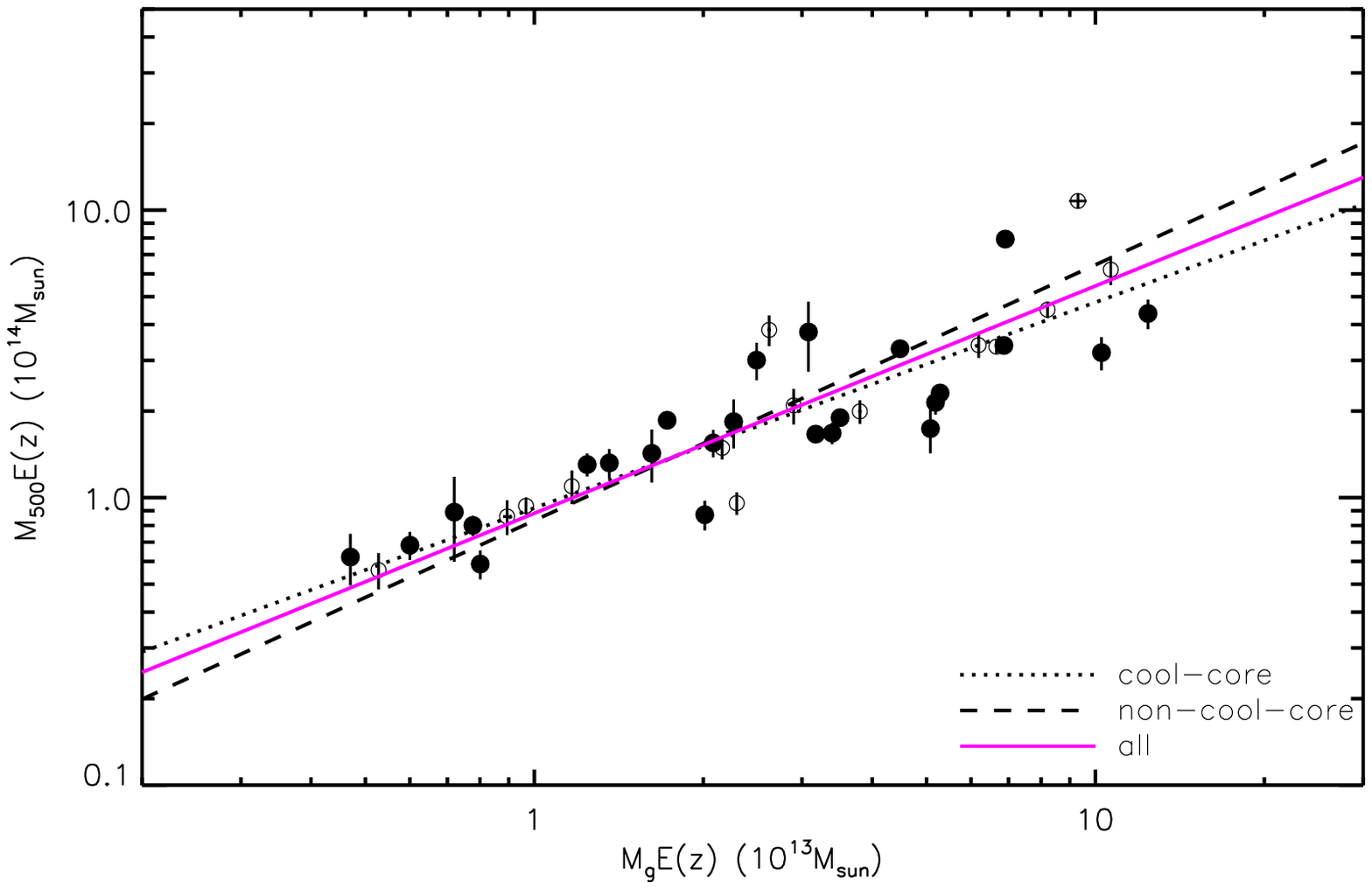,width=9.0cm,height=6.5cm}
\psfig{figure=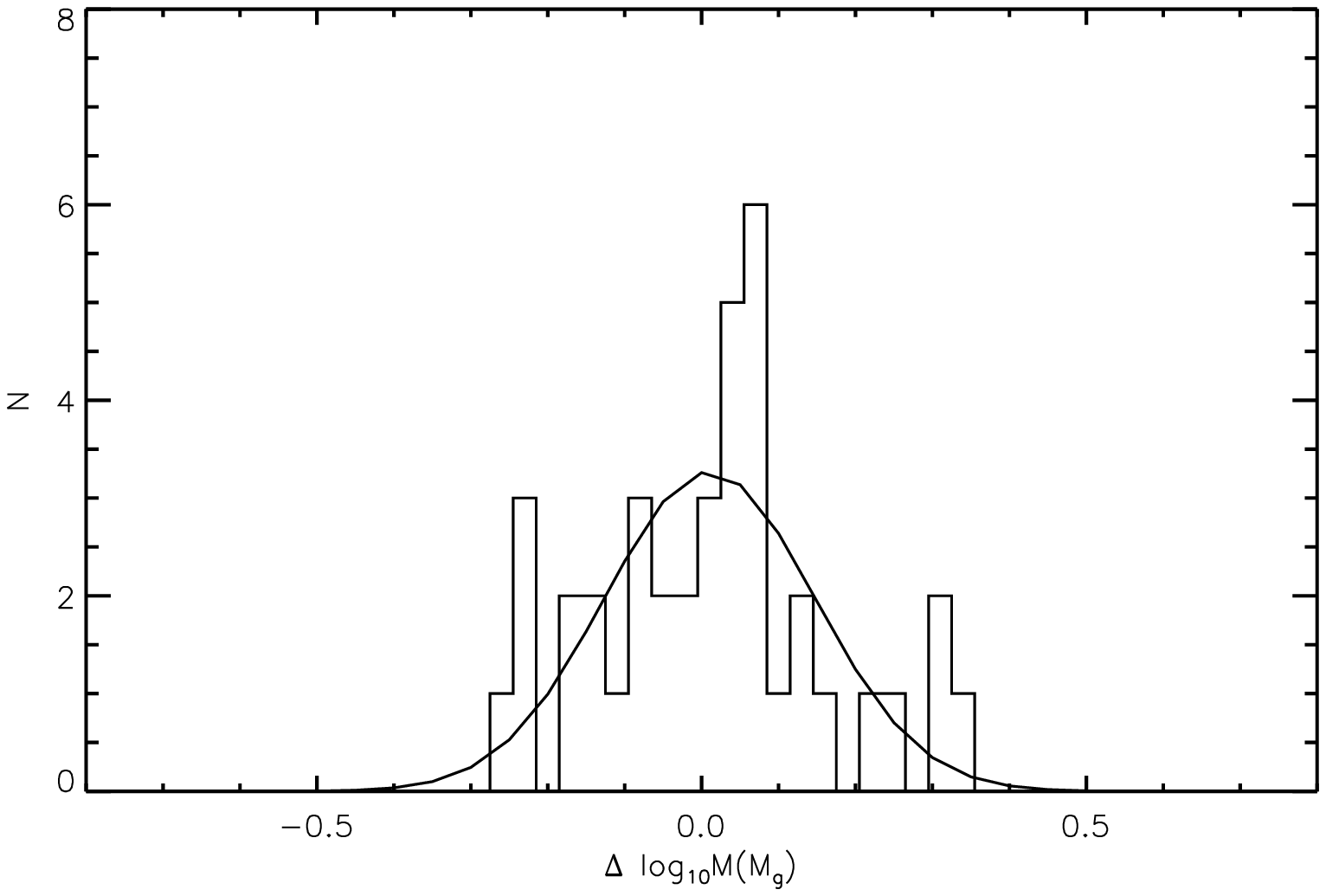,width=8.5cm,height=6.25cm}}}
\caption {Left panel: The total mass  versus gas mass inside $r_{\rm 500}$. Right panel: Histogram
of the residuals from the $M_{\rm 500}$-$M_{\rm g}$ relation in log space. This raw scatter is well
described by a Gaussian fit of $\sigma_{ M(M_{g})}$ = 0.14 dex.}
\label{zhaohh:fig5}
\end{figure}
%\clearpage

Fig.\ref{zhaohh:fig5} shows the $M_{\rm 500}$-$M_{\rm g}$ relation for our sample and the best-fit
gives a slope of $0.79 \pm 0.07$. The slope is in agreement with $0.81 \pm 0.07$ derived from {\it Chandra}
data \citep{Kravtsov06, Nagai07} and in marginal agreement with $0.91 \pm 0.08$ presented by Zhang et al. (2008).
The slope is shallower than the self-similar prediction ($M_{\rm 500}\propto {f_{\rm g}^{-1}}M_{\rm g}$).
 This may be due to the trend of observed $f_{\rm g}$ with the total mass.
Giodini et al. (2009) investigated a sample of 41 clusters, which spanned the total mass
range $1.5 \times 10^{13}\rm M_{\odot}$-$1.1 \times 10^{15} \rm M_{\odot}$, and found  $f_{g} \propto M^{0.21}$.
Taking this trend into account, the slope of the $M_{\rm 500}$-$M_{\rm g}$ relation is about 0.79, which
is in good agreement with our result. The normalization of our $M_{\rm 500}$-$M_{\rm g}$ relation is lower by $\sim 24\%$
using a fixed slope of $0.91$, and higher by $9\%$ with the slope fixed to 0.81 for the {\it Chandra} data.
The slopes of the $M_{\rm 500}$-$M_{\rm g}$ for the CCCs and NCCCs are consistent within errors.
The influence of cool core is not found on this relation.

The right panel in Fig.\ref{zhaohh:fig5} shows a histogram of the log space residuals from the $M_{\rm 500}$-$M_{\rm g}$
relation for the whole sample. This raw scatter is well described by a Gaussian fit in log space with
$\sigma_{ M(M_{g})}$ = 0.14 dex. The intrinsic scatter in $M_{\rm 500}$ around the $M_{\rm 500}$-$M_{\rm g}$
relation is 0.14 dex.

\subsection{The $L_{\rm bol}$-$M_{500}$ Relation}

\begin{figure}
\centerline{\hbox{\psfig{figure=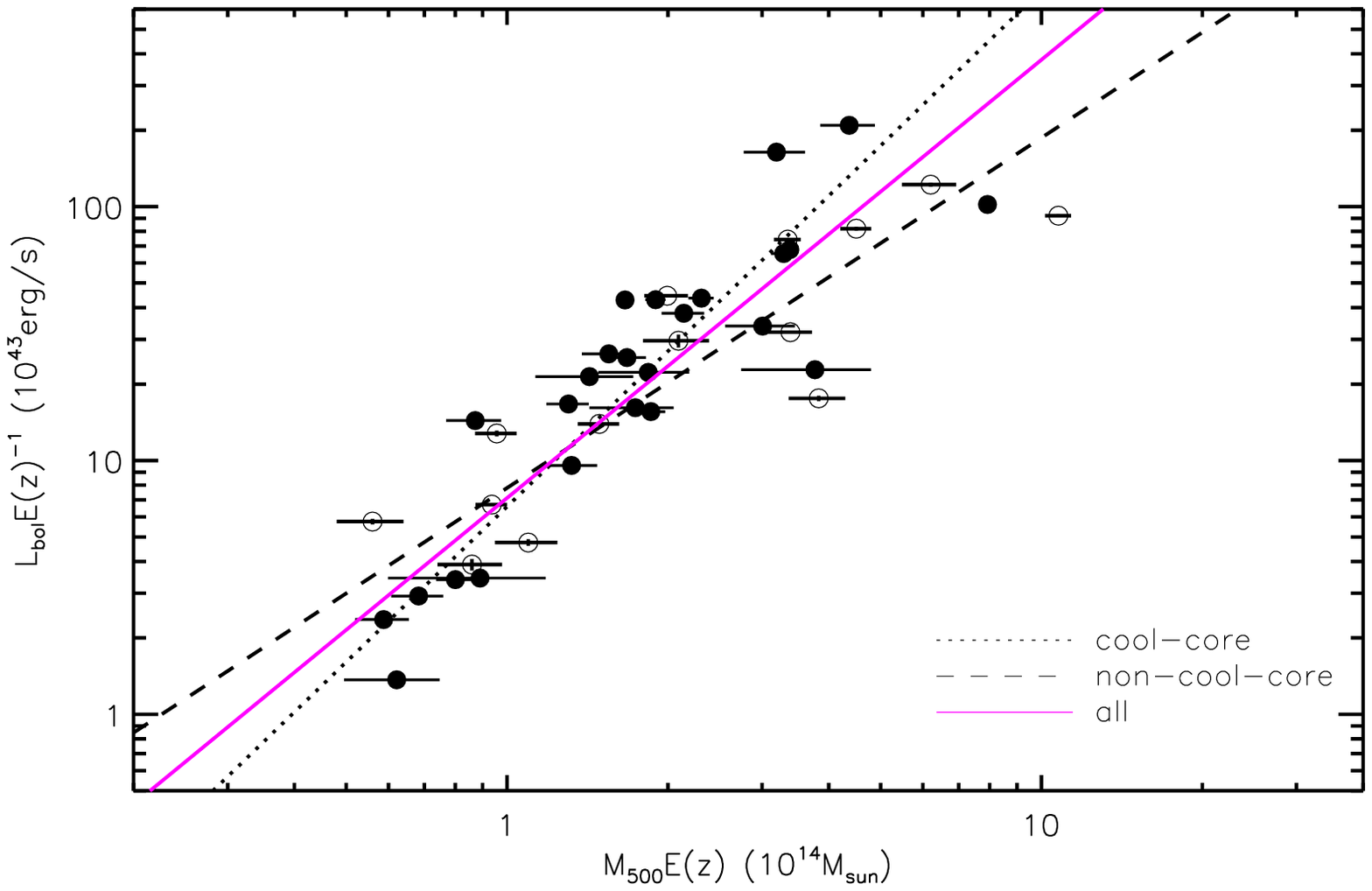,width=9.0cm,height=6.5cm}
\psfig{figure=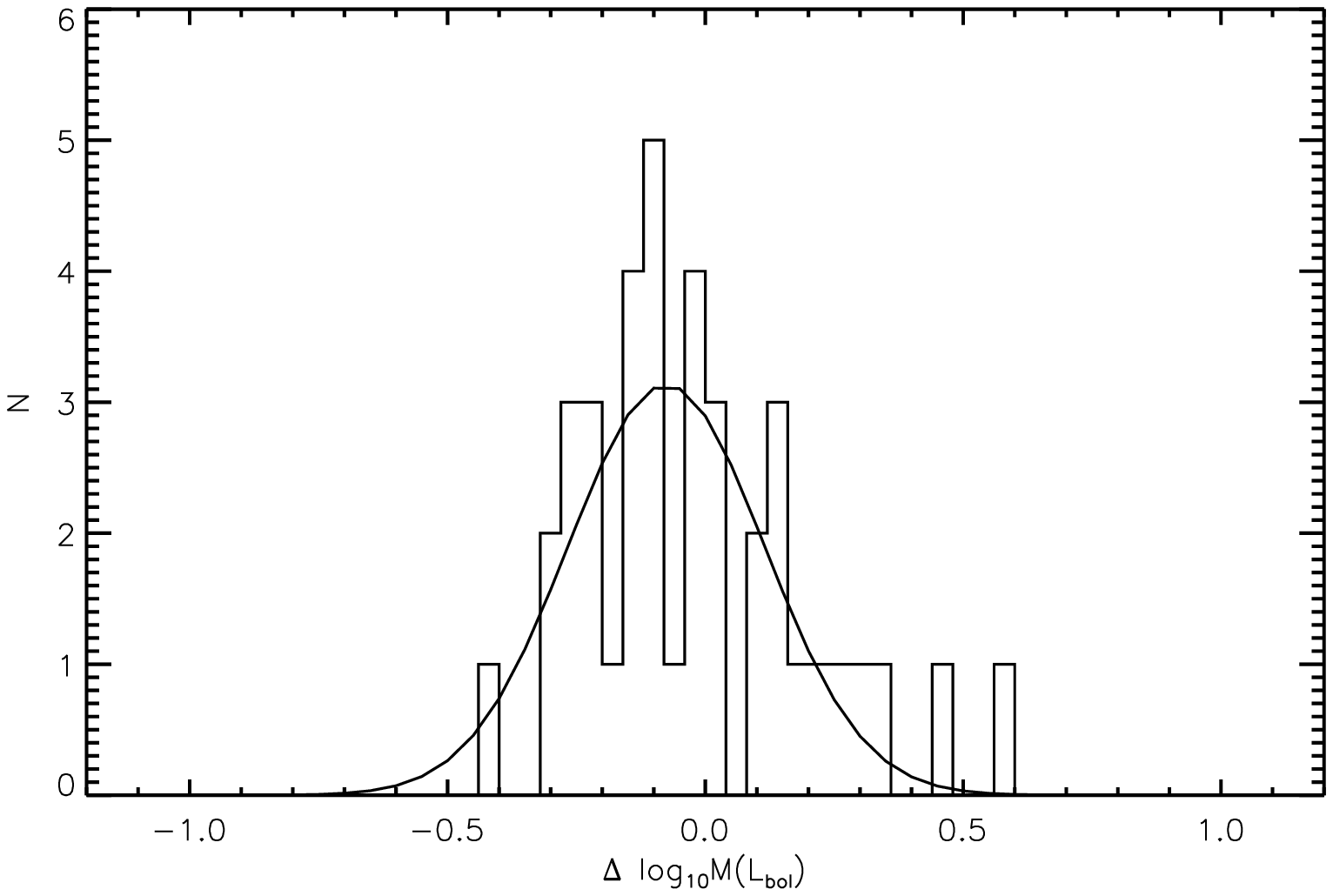,width=9.0cm,height=6.5cm}}}
\caption {Left panel: X-ray luminosity in the 0.01-100 keV  band versus the total mass.
Right panel: Histogram of the residuals from the $M_{\rm 500}-L_{\rm bol}$ relation in log space.
 This raw scatter is well described by a Gaussian fit of $\sigma_{ M(L_{bol})}$ = 0.27 dex.}
\label{zhaohh:fig6}
\end{figure}
%\clearpage
The $L_{\rm bol}$-$M_{500}$ relation is  very important  for the application to cosmological cluster
surveys. Fig.\ref{zhaohh:fig6} shows the $L_{\rm bol}$-$M_{500}$ relation for our sample and the best fit gives
a slope of $1.73 \pm 0.16$.
Our slope is higher than the self-similar expected value, 1.33, and  agrees with the results obtained
by Reiprich \& B\"ohringer (2002) ($1.80 \pm 0.08$) and Ettori et al. (2004) ($1.88 \pm 0.42$), in both of which
 the same core-uncorrected $L_{\rm bol}$ are used as ours.
Our slope is lower than  the core-corrected
result of Zhang et al. (2008, $2.36 \pm 0.07$). The slope of our
relation is consistent with the result of Morandi et al. (2007, $2.00 \pm 0.28$), in
which the $L_{\rm bol}$ was  obtained by excluding the r$<$100 kpc region.

The slope for the NCCCs ($1.38 \pm 0.17$) is consistent with the
 self-similar expected value, while the slope ($2.03 \pm 0.22$) for the CCCs is higher.
 There is a significant normalization discrepancy between CCCs and NCCCs.
 These may be due to the  high luminosity of the cool core in CCCs.

 The right panel in Fig.\ref{zhaohh:fig6} shows
 a histogram of the log space residuals from the $L_{\rm bol}$-$M_{500}$ relation
 for the whole sample. This raw scatter in $M_{500}$ is well described by a Gaussian
 fit in log space with $\sigma_{M(L_{bol})}$ = 0.27 dex. The intrinsic scatter
 in $M_{\rm 500}$ around the $L_{\rm bol}$-$M_{500}$ relation is 0.15 dex.

\subsection{The $M_{\rm 500}$-$Y_{\rm X}$ Relation}

\begin{figure}
\centerline{\hbox{\psfig{figure=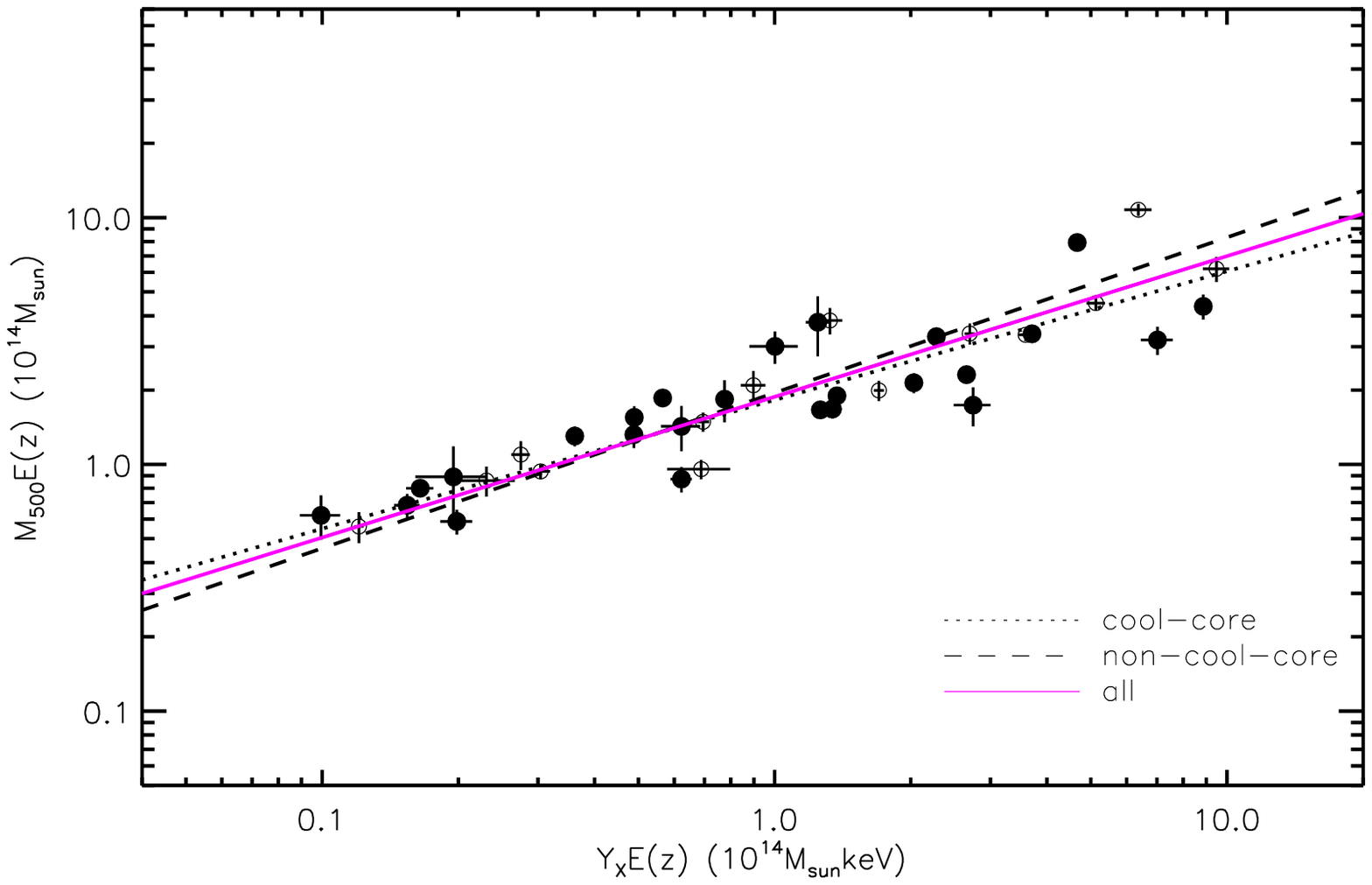,width=9.0cm,height=6.5cm}
\psfig{figure=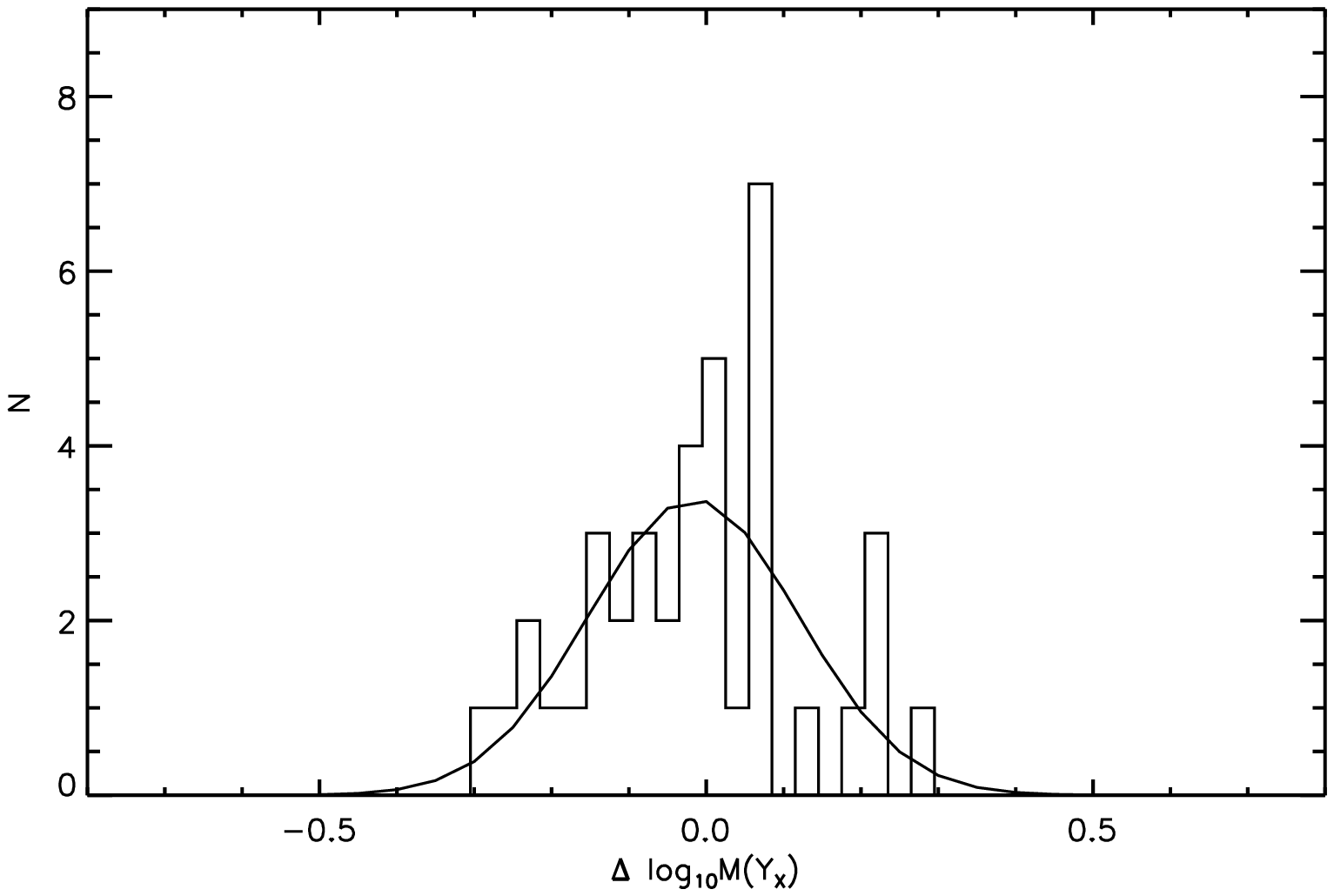,width=9.0cm,height=6.5cm}}}
\caption {Left panel: $ Y_{\rm X}$ is the product of the X-ray temperature
$T_{\rm (0.2-0.5)r_{500}}$ and gas mass $M_{\rm g}$. Right panel: Histogram
of the residuals from the $M_{\rm 500}$-$Y_{\rm X}$ relation in log space.
This raw scatter is well described by a Gaussian fit of $\sigma_{M(Y_{X})}$ = 0.13 dex.}
\label{zhaohh:fig7}
\end{figure}
%\clearpage

We present the $M_{\rm 500}$-$Y_{\rm X}$ relation for this sample in Fig.\ref{zhaohh:fig7},
in which the best-fit gives a slope of  0.57$\pm0.04$ being consistent with 0.6 expected
by the self-similar model.
Our slope is consistent with the result of $0.62 \pm 0.06$ in Zhang et al. (2008),
who used the same temperature, $T_{\rm (0.2-0.5)r_{500}}$, as ours. The slope is also in
good agreement with the value $0.57 \pm 0.03$ in Vikhlinin et al. (2009), in which
a extended temperature was used in the radial range (0.15-1)$r_{500}$ for ten
relaxed ${\it Chandra}$ clusters. Moreover, our result is supported by the
simulation ($0.57 \pm 0.01$ ) \citep{Nagai07}. Thus, the slope of
$M_{\rm 500}$-$Y_{\rm X}$ relation is stable and  not sensitive to the
temperature definition.

The right panel in Fig.\ref{zhaohh:fig7} shows a histogram of the log space
residuals from the $M_{500}$-$Y_{X}$ relation for all clusters. This raw
scatter in $M_{500}$ is well described by a Gaussian fit in log space with
$\sigma_{M(Y_{\rm X})}$ = 0.13, which is the smallest among the relation
between the observables and the total mass. We also obtain a small
intrinsic scatter  of 0.13 dex.  Fixing  the slope to 0.62,
we find that the normalization of our relation is lower than that in
Zhang et al. (2008) by  $\sim 19\%$. We also fix the slope to 0.57,
and find that the normalization of our slope is similar with
the simulation value by $\sim 2\%$.

The slope for the relation of the NCCCs agrees with that for the CCCs
within errors as listed in Table 4. Both slopes are close to the self-similar prediction (0.6).
Taking the errors into account, the normalization for the NCCCs is  also
consistent with that for the CCCs.

\section{Discussion}

\subsection{Comparison of Different Mass-Proxy Relations}
In this section, we compare the four mass scaling relations ($L_{\rm bol}$-$M_{\rm 500}$,
$M_{\rm 500}$-$T$,  $M_{\rm 500}$-$M_{\rm g}$ and  $M_{\rm 500}$-$Y_{\rm X}$),  to
find out which relation is the best to estimate cluster  mass. The comparison of these
relations are listed in Table 5.

The slope of the $L_{\rm bol}$-$M_{\rm 500}$ relation ($1.73 \pm 0.16$) is steeper
than the  self-similar expectation (1.33) significantly. It may be  due to that the
$L_{\rm bol}$ is dominated by cluster center and thus particularly
susceptible to nongravitational processes, while the self-similarity is roughly
preserved in the  outer region. For the $M_{500}$-$T$ relation, we perform
the fittings for both  the hot clusters ($T>3.0$ keV) and the whole
sample, and find that the slopes are steeper than
the self-similar value and do not  vary with the cluster temperature range. The
slope of the $M_{\rm 500}$-$M_{\rm g}$  relation is shallower than the
self-similar prediction, moreover, the gas fraction seems to depend on
the $M_{\rm 500}$ rather than  a constant. The  $M_{\rm 500}$-$Y_{\rm X}$
relation agrees with the self-similarity and is consistent with the simulations
\citep {Kravtsov05, Nagai07}.

Both  intrinsic scatters and raw scatters show that
 the $M_{\rm 500}$-$Y_{\rm X}$ relation is the tightest one compared with the $M_{\rm 500}$-$T$,
 $M_{\rm 500}$-$M_{\rm g}$ and $L_{\rm bol}$-$M_{\rm 500}$ relations.
Numerical simulations also indicate that the $Y_{\rm X}$ is a low-scatter mass proxy (with only
$\approx$ 5\%-8\% intrinsic scatter) \citep{Nagai07, Kravtsov06}, which is smaller than any other
mass proxies even in the presence of significant dynamical activity.

The $L_{\rm bol}$-$M_{\rm 500}$ relation  has the largest discrepancy in the slopes between
CCCs ($2.03 \pm 0.22$) and NCCCs ($1.38 \pm 0.17$), indicating a significant influence of
the cool core on this relation. The slope for the NCCCs is consistent with  the self-similar
expected, while the slope for the CCCs is steeper (Table 3). In contrast, the slopes of the
other  relations display much smaller differences between CCCs and NCCCs.  That indicates
the cool core has little influence on the $M_{\rm 500}$-$T$,$M_{\rm 500}$-$M_{\rm g}$ and
$M_{\rm 500}$-$Y_{\rm X}$ relations.

Many large cluster samples have been used to investigate the scaling relations.
We compare  recent results in Table 6.
Both simulations (Nagai et al. 2007, Fabjan et al. 2011) and observations (Arnaud et al. 2007,
Vikhlinin et al. 2009,  Li et al. 2013) show that the relation of $M_{\rm 500}$-$Y_{\rm X}$ is in
good agreement with the self-similar expected value, 0.6, and does not change with cluster sample,
while the other relations are quite different between samples. Compared to the other relations,
the $M_{\rm 500}$-$Y_{\rm X}$ relation is the most insensitive to variations in the physical
processes included in the simulation \citep{Pool07, Rasia11, Fabjan11}.

In sum, the $M_{\rm 500}$-$Y_{\rm X}$ relation is the stablest one among these four
scaling relations, which is in agreement with the self-similar model and does not changes
with the cluster sample. In addition, this relation  has the smallest scatter in
$M_{\rm 500}$ and does not affected by the cool core. Thus, the parameter $Y_{\rm X}$
is the best proxy to estimate cluster  mass.

\subsection{Correction for Malmquist bias}

For X-ray flux-limited cluster samples,  more luminous clusters will be selected
 from a survey volume and this will induce  the Malmquist bias.
  Many works have pointed out that the observed $L_{X}$-$M$ scaling
relation may be significantly affected by Malmquist bias if the scatter in luminosity
for fixed mass is large \citep{Stanek06, Vikhlinin09, Pratt09}. We can give the mean bias
in $\ln L$ for given mass \citep{Vikhlinin09}:
\begin{equation}
 Bias(\ln L|\ln L_0)=\langle\ln L\rangle - \ln L_0
 =
\frac
{\int_{-\infty}^{\infty}(\ln L - \ln L_0)\,p(\ln L)\,V(\ln L)\,d\ln L}
{\int_{-\infty}^{\infty} p(\ln L)\, V(\ln L)\, d\ln L}
\end{equation}
where $\ln L_0$ is the mean $L$ for given mass, and $p(\ln L)$
has a log-normal distribution ($p(\ln L)\propto-(\ln L - \ln L_0)^2/(2\sigma^2)$),
which characterizes the scatter of $L$ in the $L_{X}$-$M$ relation. For the low
redshift and flux-limited sample, the evolution on the  $L_{X}$-$M$ relation can
be neglected within the survey's effective redshift depth \citep{Vikhlinin09},
and the survey volume is a power law function of the object luminosity
($V(L)\propto L^{3/2}$ in Euclidean space). Thus, the  Eq. (11) can be worked
out analytically with $Bias(\ln L|\ln L_0)$=$3/2\sigma^2$. We also perform a Gaussian
fitting for the $L_{bol}$-$M_{500}$ relation and find
$\sigma$=0.65, the bias is 0.63. Thus, the Malmquist bias leads to a $87\%$ overestimation
in the normalization of the observed $L_{bol}$-$M_{500}$ relation in our sample, which
is much smaller than the factor of ~2 bias advocated by Stanek et al. (2006) but larger
than the value 26\% in Vikhlinin et al. (2009).

\subsection{Uncertainties in the measurement of cluster mass}

The total cluster mass may be affected by some factors.  First,
the assumption of spherical symmetry is not satisfied in some clusters, thus,
the masses of these clusters may be overestimated (or underestimated).
To estimate the uncertainty due to spherical symmetry, Landry et al. (2012)
found  a $\pm 6\%$ systematic uncertainty in the total mass of the most disturbed
clusters, A520, at $r_{\rm 500}$. All the clusters in our sample are
regular and round in  projected image, but in case of merging along the
line-of-sight, the inaccuracy  may be large due to the assumption of spherical symmetry.
Second, all the clusters in our sample are nearby, at redshift $z<0.1$.
The temperature profiles to $r_{\rm 500}$ are unavailable for  some clusters.
In this case, we use the extended temperature profile to calculate the cluster mass,
which will introduce some errors.
We select a NCCC (A3391) and a CCC (A1650), whose temperatures
at $r_{\rm 500}$ are available, to test the uncertainties due to the possibility of
mis-extrapolating the temperature profile. We fit the temperature profiles without the
outmost observed data, and then obtain their cluster masses again. Compared with the
previous masses, the biases in the cluster mass due to the mis-extrapolating temperature
profiles  for A1650 and  A3391 are ~17\% and ~4\%, respectively.
Third, the hydrostatic equilibrium may be destroyed and  non-thermal pressure
components become more significant at large radius \citep{Nagai07, Zhang08,
Vikhlinin09}. The subsonic turbulent motions of the ICM gas in relaxed clusters
in the Nagai et al. (2007) sample seem to result in a $\sim$ 15\% underestimate
in the hydrostatic estimates of $M_{\rm 500}$.

\section{Conclusion}
Using a sample of 39 X-ray  nearby ($z < 0.1$) galaxy clusters observed with {\it XMM-Newton},
we investigate the  relations between X-ray observables and total mass.
The observable parameters (e.g.  $L_{\rm bol}$, $T$, $M_{\rm g}$ and $M_{\rm 500}$) are
precisely derived by de-projecting technique.
With the criterion of the central cooling time and the central temperature,
we divide the clusters in this sample into NCCCs and CCCs and the fractions
are 36\% and 64\%, respectively.
Furthermore, we study the scaling relations of $L_{\rm bol}$-$M_{\rm 500}$,  $M_{\rm 500}$-$T$,
$M_{\rm 500}$-$M_{\rm g}$ and  $M_{\rm 500}$-$Y_{\rm X}$, and also the influences
of cool core on these relations.
The results show that the $M_{500}$-$Y_{\rm X}$ relation has a slope close to the
standard self-similar value, has the smallest scatter and does not vary
with the cluster sample. Moreover, the $M_{\rm 500}$-$Y_{\rm X}$ relation
is not affected by the cool core. Thus, the parameter of $Y_{ X}$  may be
the best  mass indicator.

\acknowledgments{} This research was supported by the National Natural Science Foundation
of China under grant Nos. 11003018, 11203019, and  by the
Strategic Priority Research Program on Space Science, the Chinese Academy of Sciences,
Grant No. XDA04010300.

\begin{table*} { \begin{center} \footnotesize
      {\renewcommand{\arraystretch}{1.3} \caption[]{
          \emph{XMM-Newton} observations and cluster properties.}
        \label{t:xmm}}
\begin{tabular}{llccccll}
\hline
Cluster&OBS-ID&$z$& $N_{\rm H}$&$T_{(0.2-0.5)r_{\rm 500}}$&$L_{\rm 0.1-2.4\mbox{} keV}$&$L_{\rm 0.01-100\mbox{} keV}$\\
       & & & $10^{20}{\rm cm}^{-2}$&keV &$10^{44}{\rm \mbox{} erg/s}$&$10^{44}{\rm \mbox{}erg/s}$  \\
\hline
2A0335  & 0109870101  & 0.0347 & 18.6 &$ 4.03 \pm 0.80 $& $2.10 \pm 0.04$&$3.44 \pm 0.07$\\
A0133   & 0144310101  & 0.0569 & 1.6  &$ 4.04 \pm 0.42 $& $1.40 \pm 0.02$&$2.34 \pm 0.06$ \\
A1650   & 0093200101  & 0.0845 & 1.5  &$ 5.49 \pm 0.37 $& $3.55 \pm 0.03$&$7.05 \pm 0.09$\\
A1795   & 0097820101  & 0.0622 & 1.2  &$ 6.75 \pm 0.23 $& $5.26 \pm 0.02$&$10.49 \pm 0.08$\\
A2029   & 0111270201  & 0.0766 & 3.2  &$ 6.70 \pm 0.48 $& $7.70 \pm 0.08$&$17.01 \pm 0.25$\\
A2052   & 0401521201  & 0.0353 & 2.9  &$ 3.16\pm 0.23 $& $0.93 \pm 0.01$&$1.46 \pm 0.02$\\
A2065   & 0202080201   & 0.0723 & 2.8  &$ 5.02 \pm 0.34 $& $2.14 \pm 0.03$&$4.40 \pm 0.13$\\
A2199   & 0008030201  & 0.0299 & 0.8  &$ 3.98 \pm 0.19 $& $2.47 \pm 0.01$&$4.35 \pm 0.03$\\
A0262   & 0109980101  & 0.0163 & 5.5  &$ 2.48 \pm 0.33 $& $0.16 \pm 0.01$&$0.24 \pm 0.01$\\
A2626   & 0148310101  & 0.0565 & 4.3  &$ 3.32 \pm 0.14 $& $1.03 \pm 0.01$&$1.60 \pm 0.03$\\
A2657   & 0505210301  & 0.0404 & 5.3  &$ 3.91 \pm 0.51 $& $1.28 \pm 0.06$&$2.18 \pm 0.21$\\
A3112   & 0105660101  & 0.0752 & 2.5  &$ 4.30 \pm 0.21 $& $3.83 \pm 0.03$&$6.77 \pm 0.08$\\
A3558   & 0107260101  & 0.0488 & 3.6  &$ 5.04 \pm 0.59 $& $0.82 \pm 0.01$&$1.65 \pm 0.04$\\
A3581   & 0205990101 & 0.023  & 4.3   &$ 2.12 \pm 0.24 $& $0.26 \pm 0.01$&$0.34 \pm 0.01$\\
A4059   & 0109950201  & 0.0475 & 1.1  &$ 3.96 \pm 0.25 $& $1.49 \pm 0.01$&$2.60 \pm 0.03$\\
A0478   & 0109880101  & 0.0882 & 15.3 &$ 7.20 \pm 0.60 $&  $10.32\pm 0.11$&$21.84 \pm 0.17$\\
A0496   & 0506260401  & 0.0326 & 5.7  &$ 3.87 \pm 0.20 $& $2.33 \pm 0.02$&$3.86 \pm 0.04$\\
S1101  & 0123900101  & 0.0564 & 1.9   &$ 2.36 \pm 0.06$&$1.93 \pm 0.01$&$2.70\pm 0.02$\\
AWM 7   & 0135950301  & 0.0172 & 9.21 &$ 3.59 \pm 0.57 $&  $0.52 \pm 0.01$&$0.96 \pm 0.10$\\
EXO0422 & 0300210401  & 0.0390 & 6.4  &$ 3.01\pm 0.18 $&$1.00 \pm 0.02$&$1.70 \pm 0.02$\\
Hercules& 0401730101  & 0.0370 & 3.4  &$ 2.70 \pm 0.78 $& $0.22 \pm 0.01$&$0.35 \pm0.01$\\
HydraA  & 0504260101  & 0.0538 & 4.86 &$ 3.52 \pm 0.11 $& $2.74 \pm 0.01$&$4.41 \pm 0.03$\\
MKW3S   & 0109930101  & 0.0442 & 3.2  &$ 3.48 \pm 0.15 $& $1.38 \pm 0.01$&$2.27 \pm 0.03$\\
MKW4    & 0093060101  & 0.0195 & 1.9  &$ 2.04 \pm 0.29 $&  $0.10 \pm 0.01$&$0.14 \pm 0.01$\\
MKW8    & 0300210701  & 0.0263 & 2.6  &$ 2.95 \pm 0.27 $&  $0.19 \pm 0.01$&$0.30 \pm 0.01$\\
%Coma   & 0153750101  & 0.0231 & 0.9   &$ 6.30 \pm 0.47$&$1.40 \pm 0.01$&$3.23 \pm 0.07$\\
%A3526   & 0406200101  & 0.0114 & 8.2  &$ 3.19 \pm 0.17 $& $1.02 \pm 0.01$&$1.87 \pm0.01$\\
%Ophiuchus& 0505150101  & 0.028 & 20.14 &$ 8.65 \pm 0.38$&$4.50 \pm 0.05$&$11.28\pm 0.09$\\
A1060  & 0206230101 & 0.0126 & 4.9    &$ 3.15 \pm 0.25$&$0.41\pm 0.01$&$0.67\pm 0.01$\\
A1651  & 0203020101  & 0.0845 & 1.7   &$ 5.42 \pm 0.31$&$3.94\pm 0.07$&$7.72 \pm 0.19$\\
A2063  & 0200120401  & 0.0358 & 2.9   &$ 3.00 \pm 0.78$&$0.89 \pm 0.03$&$1.41 \pm 0.06$\\
A2589  & 0204180101 & 0.0416 & 4.39   &$ 3.22 \pm 0.16$&$0.88 \pm 0.01$&$1.42 \pm 0.03$\\
A3158  & 0300210201  & 0.0590 & 1.1   &$ 4.40 \pm 0.18$&  $2.46 \pm 0.04$&$4.58 \pm 0.12$\\
A3391  & 0505210401  & 0.0514 & 5.4  &$ 5.44 \pm 0.74$& $0.92 \pm 0.02$&$1.80 \pm 0.07$\\
A3571  & 0086950201  & 0.0391 & 3.9  &$ 4.36 \pm 0.18$&$1.73 \pm 0.02$&$3.26 \pm 0.04$\\
A3827  & 0406200101  & 0.098  & 2.8  &$ 6.24 \pm 0.48$&   $4.71 \pm 0.06$&$8.58 \pm 0.23$\\
A0399  & 0112260301  & 0.0722 & 10.6 &$ 6.84 \pm 0.66$&$4.13 \pm 0.05$&$9.53 \pm 0.17$\\
A0400  & 0404010101  & 0.0238 & 8.9  &$ 2.29 \pm 0.21$&$0.44\pm 0.02$&$0.58\pm 0.02$\\
A4010  & 0404520501  & 0.0957 & 1.4   &$ 2.96 \pm 0.43$&$2.01 \pm 0.07$&$3.10 \pm 0.29$\\
AWM 4  & 0093060401 & 0.0326 & 4.8    &$ 2.58 \pm 0.32$&$0.25 \pm 0.01$&$0.40 \pm 0.03$\\
Triangulum& 0093620101  & 0.051 & 5.4 &$ 8.89 \pm 0.99$&$4.83 \pm 0.12$&$12.48\pm 0.36$\\
IIIZw54& 0505230401   & 0.0311 & 16.68 &$ 2.36 \pm 0.18$&$0.32 \pm 0.01$&$0.48 \pm 0.02$\\

\hline
  \end{tabular}
  \end{center}
  \hspace*{0.3cm}}
  %{\footnotesize $^{*}$ ``cc'' denotes cool core, ``S'' denotes cool-core clusters, and ``N'' denote non-cool-core clusters. }}
\end{table*}

\begin{table*} { \begin{center} \footnotesize
      {\renewcommand{\arraystretch}{1.3} \caption[]{
          \emph{XMM-Newton} observations and cluster properties.}
       \label{t:xmm}}
\begin{tabular}{lcccccc}
\hline
Cluster&$r_{500}$ & $t_{c}$ &$M_{\rm tol,500}$&$M_{\rm g}$&cc$^{*}$\\
        & Mpc  & Gyr &$10^{14} {\rm M_{\odot}}$  &$10^{13} {\rm M_{\odot}}$& \\
\hline
2A0335    &$0.93 \pm 0.01$&$0.52 \pm 0.01 $&$ 2.96 \pm 0.44  $ &$ 2.45 \pm 0.01 $& Y\\
A0133     &$1.07 \pm 0.10$&$0.87 \pm 0.01 $& $  3.67 \pm 1.00 $ &$ 3.01 \pm 0.03$&Y \\
A1650     &$1.02 \pm 0.01$&$2.17 \pm 0.07$&  $  3.25 \pm 0.13 $ &$ 6.60 \pm 0.01$&Y \\
A1795     &$1.39 \pm 0.01$&$0.84 \pm 0.01$&$  7.70 \pm 0.18 $ &$ 6.71 \pm 0.15$& Y\\
A2029     &$1.00 \pm 0.04$&$0.91 \pm 0.02$&$  3.08 \pm 0.40 $ &$ 9.89 \pm 0.04$& Y\\
A2052     &$0.66 \pm 0.03$&$0.70 \pm 0.01$&$  0.86 \pm 0.01 $ &$ 1.98 \pm 0.01$& Y\\
A2065     &$0.91 \pm 0.02$&$3.18 \pm 0.16$&$  2.29\pm 0.12 $ &$ 5.24 \pm 0.01$& Y\\
A2199     &$0.82 \pm 0.01$&$0.83 \pm 0.01$&$  1.64 \pm 0.05 $ &$ 3.13 \pm 0.01$& Y\\
A0262     &$0.59 \pm 0.02$&$0.49 \pm 0.01$&$  0.58 \pm 0.07 $ &$ 0.80 \pm 0.01$& Y\\
A2626     &$0.85 \pm 0.03$&$1.60 \pm 0.03$&$  1.81 \pm 0.14 $ &$ 1.68 \pm 0.01$& Y\\
A2657     &$0.78 \pm 0.05$&$3.68 \pm 0.95$&$  1.40 \pm 0.29 $ &$ 1.59 \pm 0.01 $& Y\\
A3112     &$1.01 \pm 0.02$&$0.76 \pm 0.01$&$  3.18 \pm 0.17 $ &$ 4.33 \pm 0.01$& Y\\
A3558     &$0.83 \pm 0.05$&$2.93 \pm 0.14$&$  1.70 \pm 0.31 $ &$ 4.97 \pm 0.01$& Y\\
A3581     &$0.65 \pm 0.02$&$0.66 \pm 0.01$&$  0.79 \pm 0.06 $ &$ 0.77 \pm 0.01$& Y\\
A4059     &$0.82 \pm 0.01$&$1.32 \pm 0.05$&$  1.64 \pm 0.14 $ &$ 3.32 \pm 0.01$& Y\\
A0478     &$1.11 \pm 0.04$&$0.81 \pm 0.01$&$  4.19 \pm 0.49 $ &$ 11.93 \pm 0.01$& Y\\
A0496     &$0.90 \pm 0.01$&$0.84\pm 0.01$&$  2.11 \pm 0.19 $ &$ 5.11 \pm 0.03$& Y\\
S1101      &$0.80 \pm 0.03$&$0.73 \pm 0.01$&$ 1.51 \pm 0.17 $ &$ 2.03 \pm 0.02$& Y \\
AWM 7     &$0.77 \pm 0.03$&$0.81 \pm 0.04$&$  1.31 \pm 0.15 $ &$ 1.35 \pm 0.01 $& Y\\
EXO0422   &$0.76 \pm 0.02$&$0.50 \pm 0.02$&$  1.28 \pm 0.12 $ &$ 1.22 \pm 0.01$& Y\\
Hercules  &$0.67 \pm 0.07$&$1.58 \pm 0.12$&$  0.88 \pm 0.29 $ &$ 0.71 \pm 0.01$& Y\\
HydraA    &$0.85 \pm 0.01$&$0.59 \pm 0.01$&$  1.85 \pm 0.08 $ &$ 3.42 \pm 0.01 $& Y\\
MKW3S     &$0.85 \pm 0.06$&$1.41 \pm 0.04$&$  1.80 \pm 0.35 $ &$ 2.22 \pm 0.01 $& Y\\
MKW4      &$0.60 \pm 0.04$&$0.57 \pm 0.01$&$  0.62 \pm 0.13 $ &$ 0.47 \pm 0.01  $& Y\\
MKW8      &$0.61 \pm 0.02$&$1.22 \pm 0.05$&$  0.68 \pm 0.08 $ &$ 0.59 \pm 0.01 $& Y\\
%Coma       &$2.33 \pm 0.03$&$21.86 \pm 2.97$ &$ 36.89 \pm 1.64 $ &$2.65 \pm 0.02 $& N\\
%A3526     &$0.75 \pm 0.01$&$0.33 \pm 0.04$&$  1.22 \pm 0.07 $ &$ 1.31 \pm 0.01$& Y\\
%Ophiuchus  &$1.81 \pm 0.01$&$1.23 \pm 0.06$&$ 17.40 \pm 0.42$ &$ 5.36 \pm 0.02 $& N\\
A1060      &$0.69 \pm 0.02$&$3.67 \pm 0.39$ &$ 0.93 \pm 0.06 $ &$ 0.96 \pm 0.01$& N\\
A1651      &$1.01\pm 0.02$&$3.80 \pm 0.16$ &$ 3.22 \pm 0.19 $  &$ 6.40 \pm 0.01 $& N \\
A2063      &$0.67 \pm 0.02$&$2.63 \pm 0.50$  &$ 0.87\pm 0.08 $ &$2.08 \pm 0.01$& N\\
A2589      &$0.79 \pm 0.02$&$2.57 \pm 0.13$ &$ 1.46 \pm 0.13 $ &$ 2.12 \pm 0.01$& N\\
A3158      &$0.86 \pm 0.03$&$7.31 \pm 0.45$ &$ 1.94\pm 0.18  $&$ 3.70 \pm 0.02$& N\\
A3391      &$1.08 \pm 0.03$&$6.79 \pm 0.12$ &$ 3.74 \pm 0.45 $&$ 2.56 \pm 0.02 $& N\\
A3571      &$1.04 \pm 0.02$&$2.42 \pm 0.14$  &$ 3.33 \pm 0.32 $ &$ 6.08 \pm 0.02$& N\\
A3827      &$1.11\pm 0.07$&$5.93 \pm 0.72$ &$  4.30 \pm 0.28  $&$ 7.84 \pm 0.01 $& N\\
A0399      &$1.51 \pm 0.03$&$7.92 \pm 0.15$ &$ 10.40 \pm 0.58 $&$ 9.00 \pm 0.32  $& N\\
A0400      &$0.58 \pm 0.03$&$1.90 \pm 0.19$  &$ 0.55 \pm 0.08 $ &$ 0.52 \pm 0.01$& N\\
A4010      &$0.89\pm 0.04$&$1.49 \pm 0.06$  &$ 2.00 \pm 0.28 $  &$ 2.77 \pm 0.04$& N\\
AWM 4      &$0.66 \pm 0.03$&$2.33 \pm 0.14$&$ 0.85 \pm 0.11 $ &$ 0.88 \pm 0.01$& N \\
Triangulum &$1.27\pm 0.05$&$9.64 \pm 0.79$  &$ 6.60 \pm 0.70 $  &$ 10.38\pm 0.02$& N\\
IIIZw54    &$0.72 \pm 0.03$&$5.11 \pm 0.62$ &$ 1.08 \pm 0.14 $ &$ 1.15 \pm 0.01$& N\\
\hline
  \end{tabular}
  \end{center}
  \hspace*{0.3cm}{\footnotesize $^{*}$ ``cc'' denotes cool core, ``Y'' denotes cool-core clusters, and ``N'' denote non-cool-core clusters. }}
\end{table*}

\begin{table}
%\begin{center}
\renewcommand{\arraystretch}{1.05}
\caption[]{Comparisons of four different fit methods for the $M_{500}$-$T$ relation. The relations are given in the form:
$\log_{10}(Y) = A + B\cdot \log_{10}(X)$.  For each method, we list the normalization parameter A, and the slop B.}
\begin{tabular}{llc}
\hline
\hline
   Method     & $B$               & $A$\\
\hline
BCES(Y$|$X)          & $    2.22 \pm 0.42  $ & $13.85\pm 0.13$ \\
BCES(X$|$Y)          & $    1.94 \pm 0.18  $ & $13.28\pm 0.07$ \\
BCES Bisector        & $    1.94 \pm 0.17  $ & $13.12\pm 0.01$\\
BCES Orthogonal      & $    1.99 \pm 0.20  $ & $13.15\pm 0.08$ \\
\hline
\end{tabular}
%\end{center}
\end{table}

\begin{table*}
\renewcommand{\arraystretch}{1.05}
\begin{center}
\caption[]{Summary of the fits to the mass scaling relations. The relations are given in the form:
$\log_{10}(Y) = A + B\cdot \log_{10}(X)$. For each relation, we list the normalization parameter ${A}$, and the slop B.
  In addition to the results for all the sample,  we also list the results both for  the CCC sub-sample, and the NCCC sub-sample.}
%\begin{tabular}{c|ccc|ccc}
\begin{tabular}{ccccccc}
\hline
\hline
 $Y$ &$X$ & Number of clusters & $B$ & ${A}$  & comments  \\
\hline
$\frac{M_{\rm 500}}{\rm M_{\odot}} \; E(z) $
&$\frac{T}{\rm keV}$
    & 39    & $1.94 \pm{0.17}$ & ${13.12 \pm 0.10}$ & ALL  \\
&   & 25    & $1.85 \pm{0.22}$ & ${13.17 \pm 0.12}$ & CCCs \\
&   & 14    & $2.03 \pm{0.26}$ & ${13.08 \pm 0.15}$ & NCCCs \\
\hline
$\frac{M_{\rm 500}}{\rm M_{\odot}}\;E(z) $
&$\frac{M_{\rm g}}{\rm M_{\odot}} \; E(z) $
    & 39    & $0.79 \pm{0.07}$ & ${3.67 \pm 0.95}$ & ALL  \\
&   & 25    & $0.72 \pm{0.09}$ & ${4.66 \pm 1.17}$ & CCCs \\
&   & 14    & $0.89 \pm{0.11}$ & ${2.36 \pm 1.41}$ & NCCCs \\
\hline
$\frac{L_{\rm bol}}{\rm erg\;s^{-1}}\;E(z)^{-1} $
& $\frac{M_{\rm 500}}{\rm M_{\odot}} \; E(z) $
    & 39    & $1.73 \pm{0.16}$ & ${19.68 \pm 2.26}$  & ALL  \\
&   & 25    & $2.03 \pm{0.22}$ & ${15.35 \pm 3.19}$  & CCCS \\
&   & 14    & $1.38 \pm{0.17}$ & ${24.58 \pm 2.33}$ & NCCCs \\
\hline
$\frac{M_{\rm 500}}{\rm M_{\odot}}\;E(z) $
&$\frac{Y_{\rm X}}{\rm M_{ \odot}\;{\rm keV}} \; E(z) $
  & 39    & $0.57 \pm{0.04}$ & ${6.29 \pm 0.58}$ & ALL  \\
& & 25    & $0.52 \pm{0.06}$ & ${5.47 \pm 0.80}$ & CCCs \\
& & 14    & $0.63 \pm{0.06}$ & ${6.95 \pm 0.76}$ & NCCCs \\
\hline
\end{tabular}
\end{center}
\end{table*}

\begin{table*}
\renewcommand{\arraystretch}{1.05}
%\begin{center}
\caption[]{Comparison of the mass relations in our cluster sample. We  list the results  for  the whole sample.
 The raw scatters are derived by a Gaussian fit  to the  residual histograms. The intrinsic scatters are obtained from Eq. (10).}
%\begin{tabular}{c|ccc|ccc}
\begin{tabular}{lccccc}
\hline
\hline
 $$  & self-similarity & $B$ & ${A}$ & scatter (dex)& intrinsic scatter (dex)\\
\hline
$M_{\rm 500}E(z)$-$T$
    & 1.5    & $1.94 \pm{0.17}$ & ${13.12 \pm 0.10}$  &0.18 &0.14\\
%\hline
$L_{\rm bol}E(z)^{-1}$-$M_{\rm 500}E(z) $
    & 1.33    & $1.73 \pm{0.16}$ & ${19.68 \pm 2.26}$ & 0.27&0.15  \\
%\hline
$M_{\rm 500}E(z)$-$M_{\rm g}E(z) $
    & 1.0    & $0.79 \pm{0.07}$ & ${3.67 \pm 0.95}$ &0.14&0.14 \\
%\hline
$M_{\rm 500}E(z)$-$Y_{\rm X}E(z) $
  & 0.6    & $0.57 \pm{0.04}$ & ${6.29 \pm 0.58}$ &0.13&0.13 \\
\hline
\end{tabular}
%\end{center}
\end{table*}

%---------------------------------------------------
   \begin{table}
      \caption{Recent results on the slopes of the scaling relations.}
%         \label{Tempx}
      \[
         \begin{array}{llll}
            \hline
            \hline
            \noalign{\smallskip}
{\rm relation} & {\rm slope} & {\rm comments}  & {\rm reference}\\
            \noalign{\smallskip}
            \hline
 L_{bol}-M_{500}     & 1.63 \pm 0.08  &  {\rm 115~clusters,~{\rm z} = 0.1-1.3,~core~ excised~L_{bol},~r<0.15r_{500}} & {\rm Maughan07}\\
% L_{bol} - M_{500}     & 1.63 \pm 0.08  &   {\rm Core~excised~luminosity}  & {\rm Maughan07}\\
 L_{0.1-2.4} - M_{500}      & 1.82 \pm 0.13  & {\rm 106~clusters,}~\it ROSAT/ASCA   & {\rm Chen07}\\
 L_{bol} - M_{500}      & 1.71 \pm 0.46  &  {\rm 24~clusters,}~{\it Chandra},~core~ excised~L_{bol},~r<100\rm kpc   & {\rm Morandi07}\\
 L_{bol} - M_{500}     & 2.33 \pm 0.70  &  {\rm 37~clusters,}~{\it XMM},~core~ corrected ~L_{bol},~r<0.2r_{500}  & {\rm Zhang08}\\
 L_{0.1-2.4} - M_{500} & 1.76 \pm 0.13  &  {\rm 31~clusters,}~{\it XMM} & {\rm Arnaud10}\\
 L_{bol} - M_{500}     & 1.51 \pm 0.09  & {\rm 14~literature~samples}    & {\rm Reichert11}\\
            \noalign{\smallskip}
            \hline
            \noalign{\smallskip}
 M_{500} - T    &  1.54 \pm 0.06    & {\rm 106~clusters,}~\it ROSAT/ASCA  & {\rm Chen07}\\
 M_{500} - T    &  1.71 \pm 0.09    &  {\rm  10~clusters,}~T_{0.1-0.5r_{200}},~\it XMM  & {\rm Arnaud07} \\
 M_{500} - T    &  1.74 \pm 0.09    &  {\rm 70~clusters,}~z=0.18-1.24  & {\rm O'Hara07} \\
 M_{500} - T   & 1.65 \pm 0.26    &  T_{0.2-0.5r_{500}}  & {\rm Zhang08}\\
 M_{500} - T   & 1.53 \pm 0.08    &  {\rm 17~clusters,}~T_{0.15-1r_{500}},~{\it Chandra}  &{\rm Vikhlinin09}\\
 M_{500} - T   & 2.04 \pm 0.04    &  {\rm 238~clusters,}~{\it Chandra/ROSAT} & {\rm Mantz10}\\
 M_{500} - T   & 1.76 \pm 0.08      & {\rm 14~literature~samples}  & {\rm Reichert11} \\
 \noalign{\smallskip}
            \hline
            \noalign{\smallskip}
 M_{500} -  M_{g}    &  0.80 \pm 0.04    &{\rm 10~clusters,}~{\it XMM}& {\rm Arnaud07}\\
 M_{500} - M_{g}     &  0.81 \pm 0.07    &  {\rm simulation}  & {\rm Nagai07} \\
 M_{500} - M_{g}    & 0.91 \pm 0.08    & {\rm 37~clusters,}~{\it XMM} & {\rm Zhang08}\\
            \noalign{\smallskip}
            \hline
            \noalign{\smallskip}
 M_{500} - Y    &  0.57 \pm 0.01    &  {\rm simulation}& {\rm Nagai07}\\
 M_{500} - Y    &  0.59 \pm 0.01    &  {\rm simulation}  & {\rm Fabjan11} \\
 M_{500} - Y   & 0.56 \pm 0.03    &{\rm 10~clusters,}~{\it XMM}& {\rm Arnaud07}\\
 M_{500} - Y   & 0.57 \pm 0.03    &{\rm 17~clusters,}~{\it Chandra}&{\rm Vikhlinin09}\\
 M_{500} - Y   &0.62 \pm 0.06    & {\rm 37~clusters,}~{\it XMM}& {\rm Zhang08}\\
%    \hline
            \hline
         \end{array}
      \]
{The references are from top to bottom:
Maughan et al. (2007), Chen et al.  (2007), Morandi et al. (2007), Zhang et al. (2008),
Arnaud et al. (2010), Reichert et al. (2011), Arnaud et al. (2007), O'Hara et al. (2007),
Vikhlinin et al. (2009), Mantz et al. (2010), Nagai et al. (2007), Fabjan et al. (2011).}

   \end{table}
%-------------------------------------------------

\end{document}